\documentclass[a4paper,fleqn,usenatbib]{mnras}

\usepackage{savesym}
\usepackage{amsmath}
\savesymbol{iint}
\savesymbol{iiint}
\usepackage{txfonts}
\restoresymbol{TXF}{iint}
\restoresymbol{TXF}{iiint}

\usepackage{graphicx}	
\usepackage{arydshln}	

\newcommand{\SKIRT}{{\large\scshape skirt}}
\newcommand{\Msun}{\mathrm{M}_\odot}
\newcommand{\Zsun}{\mathrm{Z}_\odot}
\newcommand{\HI}{\ion{H}{I}}
\newcommand{\HII}{\ion{H}{II}}
\newcommand{\C}{$\mathcal{C}$}
\newcommand{\F}{$\mathcal{F}$}

\newcommand{\ch}[1]{{#1}}   

\title[FIR and dust properties of EAGLE galaxies]{Far-infrared and dust properties of present-day galaxies in the EAGLE simulations}

\author[P. Camps et al.]{%
Peter Camps,$^{1}$\thanks{E-mail: peter.camps@ugent.be}
James W. Trayford,$^{2}$
Maarten Baes,$^{1}$
Tom Theuns,$^{2}$\newauthor
Matthieu Schaller,$^{2}$
Joop Schaye$^{3}$
\\
\\
$^{1}$Sterrenkundig Observatorium, Universiteit Gent, Krijgslaan 281, B-9000 Gent, Belgium\\
$^{2}$Institute for Computational Cosmology, Department of Physics, University of Durham, South Road, Durham DH1 3LE, UK\\
$^{3}$Leiden Observatory, Leiden University, PO Box 9513, NL-2300 RA Leiden, the Netherlands
}

\date{Accepted XXX. Received YYY; in original form ZZZ}

\pubyear{2016}

\begin{document}
\label{firstpage}
\pagerange{\pageref{firstpage}--\pageref{lastpage}}
\maketitle

\begin{abstract}
The EAGLE cosmological simulations reproduce the observed galaxy stellar mass function and many galaxy properties. In this work, we study the dust-related properties of present-day EAGLE galaxies through mock observations in the far-infrared and submm wavelength ranges obtained with the 3D dust radiative transfer code \SKIRT. To prepare an EAGLE galaxy for radiative transfer processing, we derive a diffuse dust distribution from the gas particles and we re-sample the star-forming gas particles and the youngest star particles into star-forming regions that are assigned dedicated emission templates. We select a set of redshift-zero EAGLE galaxies that matches the $K$-band luminosity distribution of the galaxies in the \emph{Herschel} Reference Survey (HRS), a volume-limited sample of about 300 normal galaxies in the Local Universe. We find overall agreement of the EAGLE dust scaling relations with those observed in the HRS, such as the dust-to-stellar mass ratio versus stellar mass and versus $\mathrm{NUV}-r$ colour relations. A discrepancy in the $f_{250}/f_{350}$ versus $f_{350}/f_{500}$ submm colour-colour relation implies that part of the simulated dust is insufficiently heated, likely because of limitations in our sub-grid model for star-forming regions. We also investigate the effect of adjusting the metal-to-dust ratio and the covering factor of the photodissociation regions surrounding the star-forming cores. We are able to constrain the important dust-related parameters in our method, informing the calculation of dust attenuation for EAGLE galaxies in the UV and optical domain.
\end{abstract}



\section{Introduction}

Cosmological simulations are a valuable tool in the study of how galaxies form and evolve. Recently, hydrodynamical simulations of the formation of galaxies in cosmologically representative volumes have succeeded in reproducing many -- but not all -- observed properties of galaxies and of the intergalactic medium to unprecedented levels of agreement \citep[e.g.,][]{LeBrun2014, Vogelsberger2014, Schaye2015}. The mass resolution for baryonic matter in these simulations is on the order of $10^6$ solar masses. Physical processes on unresolved scales (including star formation and stellar feedback) are handled through sub-grid prescriptions. Zoom-in simulations \citep[e.g.,][]{Hopkins2014, Wang2015, McKinnon2016, Sawala2016} offer a better resolution, however, they still use similar sub-grid prescriptions. Inevitably these limitations lead to uncertainties in some of the simulation predictions.

By comparing simulation results and observations we hope to examine the empirical scaling laws, deduce improved sub-grid prescriptions, and eventually, to further our understanding of the underlying physical processes. Because properties of real galaxies are derived from observed quantities (i.e.\ fluxes), they may be subject to unknown systematic biases. Making mock observations of simulated galaxies enables direct comparison to observational data, and helps to characterise the systematics involved in the transformation between intrinsic and observed quantities \citep[see, e.g.,][]{Hayward2015, Guidi2015}.

Extinction by dust grains residing in the interstellar medium (ISM) can substantially influence the flux detected from a galaxy in the UV and optical wavelength ranges. It is very hard to estimate the dust mass in a galaxy based solely on the information at these wavelengths, and thus it is difficult to account accurately for the dust obscuration effect \citep[e.g.,][]{Disney1989, Byun1994}. To alleviate this limitation, one can turn to the far-infrared (FIR) to submm wavelength range. In this window, the continuum spectra of star-forming galaxies are dominated by thermal emission from dust grains that reprocess the UV/optical radiation, providing an independent and more direct measurement of the amount of dust in a galaxy. This additional information is especially useful for constraining the dust modelling of numerically simulated galaxies that have no explicit dust component. On the other hand, accurately predicting dust emission from a simulated galaxy requires solving a nontrivial 3D radiative transfer problem \citep[see, e.g.,][]{Whitney2011, Steinacker2013}.

\ch{Several authors have performed UV to submm radiative transfer modelling for up to a few dozen simulations of isolated galaxies or galaxy mergers \citep[e.g.,][]{Narayanan2010a, Narayanan2010b, Jonsson2010, Scannapieco2010, Hayward2011, Hayward2012, Hayward2013, Hayward2014, Hayward2015, Robitaille2012, Dominguez-Tenreiro2014, Saftly2015}. While \citet{Torrey2015} do produce mock images and SEDs for thousands of present-day galaxies in the cosmological simulation Illustris \citep{Vogelsberger2014}, they do not include dust emission, limiting the observables to UV, optical and near-infrared (NIR) wavelengths. In this present work, we study the effects of dust in the full UV to submm wavelength range. We use simulated galaxies that were evolved as part of a cosmologically representative volume, and that are available in sufficiently large numbers to allow a statistically relevant confrontation with observations.}

\ch{Specifically}, we concentrate on the FIR and dust-related properties of the present-day galaxies produced by the EAGLE simulations \citep{Schaye2015, Crain2015}. EAGLE is a suite of hydrodynamical simulations of the formation of galaxies in cosmologically representative volumes, with sub-grid models for radiative cooling, star formation, stellar mass loss, and feedback from stars and accreting black holes. The sub-grid physics recipes are calibrated to reproduce the present-day galaxy stellar mass function and galaxy sizes, and show good agreement with many observables not considered in the calibration, including present-day specific star-formation rates, passive fractions, the Tully-Fisher relation \citep{Schaye2015}, and the neutral gas content \citep{Bahe2016}. The simulations also track the observed evolution of the galaxy stellar mass function out to redshift $z = 7$ \citep{Furlong2015} and reproduce the observed optical colours for galaxies in the Local Universe \citep{Trayford2015, Trayford2016}.

We use the \emph{Herschel} Reference Survey \citep{Boselli2010} (HRS), a volume-limited sample of about 300 `normal' galaxies in the Local Universe, as a reference for observed dust properties. We select a set of redshift-zero EAGLE galaxies that matches the $K$-band luminosity distribution of the HRS galaxies, and we use the 3D dust radiative transfer code \SKIRT\ \citep{Baes2011, Camps2015a} to calculate observable properties for these galaxies from UV to submm wavelengths. We compare the stellar mass, dust mass, and star-formation rate derived from our mock observations through standard tracers with the intrinsic EAGLE values, and we compare the EAGLE dust scaling relations with those observed for HRS galaxies presented by \citet{Boselli2012} and \citet{Cortese2012}. Finally, we investigate the effect of varying dust-related parameters in our post-processing procedure. This allows us to constrain these parameters, thus informing the calculation of dust attenuation for EAGLE galaxies in the UV and optical domain by \citet{Trayford2016}.

In Sect.~\ref{Methods.sec} we provide some background on the EAGLE simulations and the \SKIRT\ radiative transfer code, and we describe how the EAGLE results were exported to and post-processed by \SKIRT, with some details relayed to the appendices. In Sect.~\ref{Results.sec} we present and discuss the results of our analysis, and in Sect.~\ref{Conclusions.sec} we summarise and conclude.


\section{Methods}
\label{Methods.sec}

\subsection{The EAGLE simulations}
\label{EAGLE.sec}


\begin{table}
\caption{We use the redshift-zero snapshots of the three EAGLE simulations listed in this table. We refer to them through the labels in the first column. The second column shows the corresponding full EAGLE simulation name as defined in Tables 2 and 3 in \citet{Schaye2015}. The remaining columns list the simulation's co-moving box size, the initial number of baryonic particles, the initial baryonic particle mass, and the maximum proper gravitational softening length (i.e.\ at redshift zero). }
\label{eaglesims.tab}
{\renewcommand{\arraystretch}{1.15} \setlength{\tabcolsep}{4.5pt}
\begin{tabular}{l l r r r r}
\hline
Label  &  EAGLE name  &  $L$  &  $N$  &  $m_\mathrm{g}$  &  $\epsilon_\mathrm{prop}$  \\
           &                         & (cMpc) &          &    $(\Msun)$      &    (kpc)  \\
\hline
Ref100  & Ref-L100N1504   &  $100$ & $1504^3$ & $1.81\times10^6$ & $0.70$ \\
Recal25 & Recal-L025N0752 &   $25$ &  $752^3$ & $2.26\times10^5$ & $0.35$ \\
Ref25   & Ref-L025N0752   &   $25$ &  $752^3$ & $2.26\times10^5$ & $0.35$ \\
\hline
\end{tabular}}
\end{table}


The Evolution and Assembly of GaLaxies and their Environments (EAGLE) project \citep{Schaye2015, Crain2015} is comprised of a suite of smoothed particle hydrodynamics (SPH) simulations that follow the formation of galaxies and large-scale structure in cosmologically representative volumes of a standard $\Lambda$ cold dark matter universe. EAGLE uses the hydrodynamics code {\large\scshape gadget} \citep[first described by][]{Springel2005b}, but employs an improved hydrodynamics scheme, referred to as {\large\scshape anarchy}, described by \citet{Schaye2015} and \citet{Schaller2015}. The sub-grid models used in EAGLE are based on those developed for OWLS \citep{Schaye2010}. They are described in detail in \citet{Schaye2015} and summarised very briefly below.

Hydrogen reionisation is modelled by turning on the time-dependent, spatially uniform UV/X-ray background from \citet{Haardt2001} at redshift $z = 11.5$. Radiative cooling and photo-heating are implemented element by element following \citet{Wiersma2009a}, including all 11 elements that they found to dominate the radiative rates. Star formation follows \citet{Schaye2008}, but with the metallicity-dependent density threshold of \citet{Schaye2004}. Stellar mass-loss and chemical enrichment is based on \citet{Wiersma2009b} and tracks the elements H, He, C, N, O, Ne, Mg, Si, and Fe individually, while fixed abundance ratios relative to Si are assumed for Ca and S. Energetic feedback from star formation uses a stochastic thermal feedback scheme following \citet{DallaVecchia2012}, with a variable efficiency depending on local gas density and metallicity. A super-massive black hole seed is placed at the centre of every halo above a threshold mass \citep{Springel2005a} and is allowed to grow through gas accretion and mergers \citep{Rosas-Guevara2015, Schaye2015}. Feedback from these accreting black holes quenches star formation in massive galaxies, shapes the gas profiles in the inner parts of their host halos, and regulates the growth of the black holes themselves.

A drawback for the purpose of this work is that the EAGLE simulations do not model the cold gas phase in the ISM \citep[see Sect.~4.3 of ][]{Schaye2015}. To limit the pressure of star-forming gas particles, the EAGLE simulations impose a temperature floor, $T_\mathrm{eos}(\rho)$, as a function of the local gas density, $\rho$, corresponding to the polytropic equation of state $\rho\,T_\mathrm{eos}\propto P_\mathrm{eos}\propto\rho^{4/3}$ \citep{Schaye2008}. As a consequence, there are no resolved molecular clouds. Instead, the simulated ISM consists of smoothly distributed, warm gas. We address this issue to some extent by employing a separate sub-grid model for star-forming regions in our post-processing procedure (see Sect.~\ref{SEDs.sec}), and by assigning dust to star-forming gas particles regardless of their imposed, unphysical temperature (see Sect.~\ref{DustFraction.sec}). It remains important, however, to keep this limitation in mind when interpreting our results.

To enable numerical convergence studies, the EAGLE suite includes simulations with varying spatial resolution and simulation volume. In this work, we use the redshift-zero snapshots of the three EAGLE simulations listed in Table~\ref{eaglesims.tab}. The sub-grid prescriptions in the EAGLE reference simulation (`Ref100' in Table~\ref{eaglesims.tab}) are calibrated to reproduce the present-day galaxy stellar mass function. One of the higher-resolution simulations (`Ref25' in Table~\ref{eaglesims.tab}) employs the same sub-grid parameter values, i.e.\ calibrated for the resolution of the Ref100 simulation. For the other simulation (`Recal25' in Table~\ref{eaglesims.tab}), the sub-grid prescriptions have been re-calibrated to compensate for the effects of the increased numerical resolution. This approach allows investigating the `weak' and `strong' convergence properties of the simulations, as explained in \citet{Schaye2015}.

The public database presented by \citet{McAlpine2016} lists a wide range of properties for the galaxies in the EAGLE simulations, including intrinsic quantities obtained by integrating over particle properties, luminosities in various optical and near-infrared bands (ignoring extinction by dust), and mock optical thumbnail images. When referring to a specific galaxy in this work, we specify the unique identifier (`GalaxyID') associated with that galaxy in the public EAGLE database.

\subsection{Galaxy selections}
\label{GalaxySelections.sec}


\begin{figure}
  \includegraphics[width=\columnwidth]{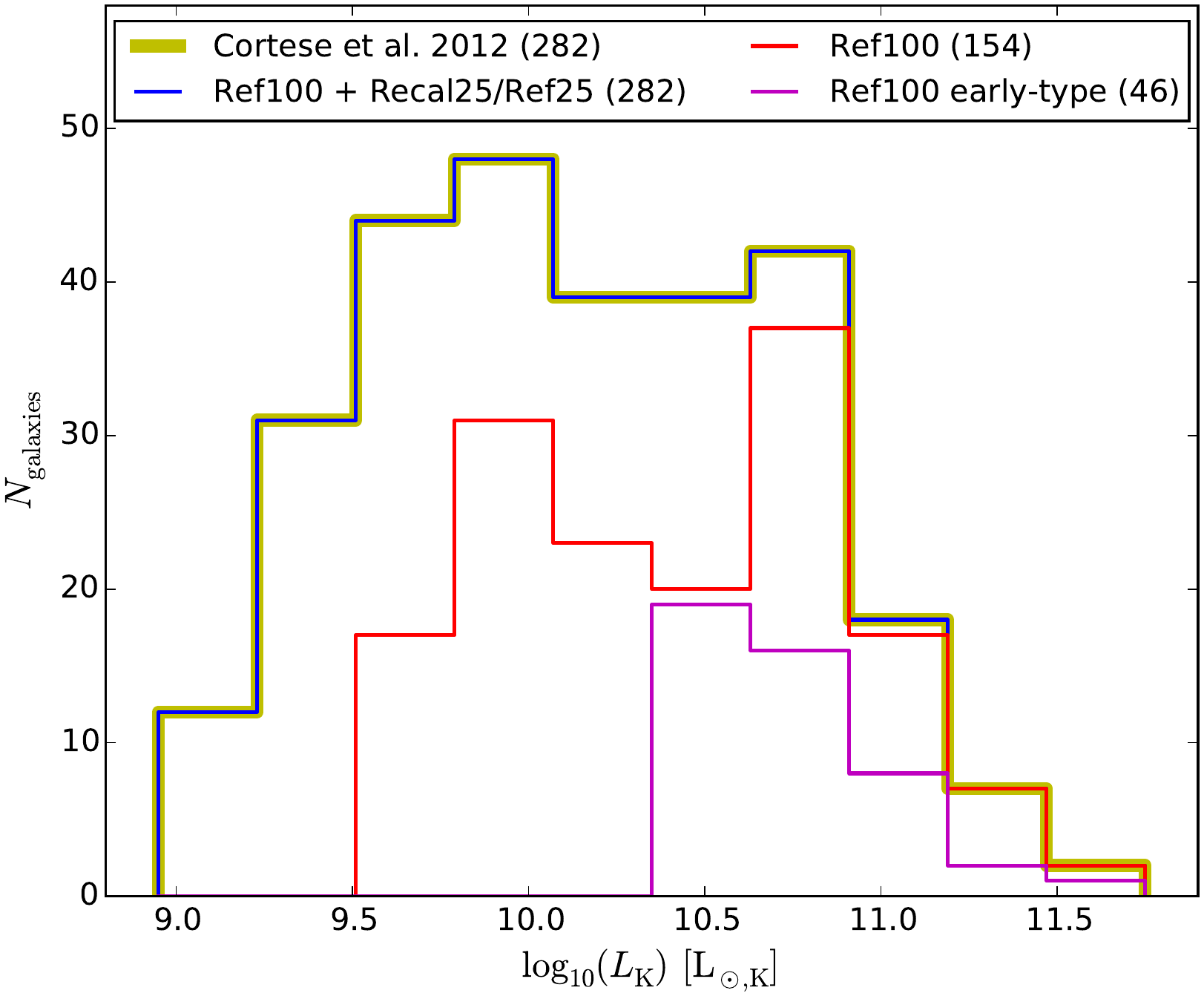}
  \caption{The $K$-band luminosity distribution of the galaxies in the \citet{Cortese2012} HRS sub-sample (dark yellow) and of the EAGLE galaxies selected for this work (blue) to match that sample. The curves are identical for both of the sets \C\ and \F\ listed in Table~\ref{GalaxySets.tab}. The distribution of the galaxies taken from the standard-resolution snapshot Ref100 is shown in red; the remainder of the galaxies are taken from one of the higher-resolution snapshots, i.e.\  either Recal25 or Ref25. The distribution of the early-type galaxies, which are all taken from the Ref100 snapshot, is shown in magenta.}
  \label{KBandHistogram.fig}
\end{figure}


\subsubsection{The HRS galaxies}
\label{HRS.sec}

In Sect.~\ref{Results.sec} we compare the dust-related properties of EAGLE galaxies with the observed properties of the galaxies in the \emph{Herschel} Reference Survey \citep[HRS,][]{Boselli2010}, and more specifically the subset presented by \citet{Cortese2012}. 

The HRS consists of a volume-limited sample ($15 \leqslant D \leqslant 25~\mathrm{Mpc}$) including late-type galaxies with 2MASS \citep{Skrutskie2006} $K$-band magnitude $K_\mathrm{S}\leqslant 12~\mathrm{mag}$ and early-type galaxies with $K_\mathrm{S}\leqslant 8.7~\mathrm{mag}$. The total sample consists of 322 galaxies (260 late- and 62 early-type galaxies). As argued by \citet{Boselli2010}, this sample is representative of the Local Universe and it spans different density regimes from isolated galaxies to the center of the Virgo cluster.

The HRS sub-sample analysed by \citet{Cortese2012} includes only those galaxies for which \emph{Herschel} as well as H\textsc{i}, NUV and SDSS observations are available, i.e.\ a total of 282 galaxies (234 late- and 48 early-type galaxies). As argued by \citet{Cortese2012}, the sub-sample is representative of the full HRS sample, and it is thus representative of the local galaxy population as well.

According to \citet{Hughes2013} and \citet{Viaene2016a}, only 5 to 8 per cent of the HRS galaxies potentially host an active galactic nucleus (AGN), depending on the criteria used. Furthermore, \citet{Viaene2016a} argue that the dust attenuation properties of the potential AGN hosts (and thus their FIR emission) do not differ fundamentally from those of the other galaxies in the sample. Consequently, we do not exclude or single out these galaxies.


\begin{table}
\caption{Characteristics of the two sets of EAGLE galaxies for which we present results in this work. The first two columns show a symbol to identify the set and a mnemonic for the origin of this symbol. Subsequent columns list the total number of galaxies in the set and the number of galaxies extracted from each of the EAGLE snapshots used in this work (see Table~\ref{eaglesims.tab}). The final column shows the number of early-type galaxies in each set.}
\label{GalaxySets.tab}
{\renewcommand{\arraystretch}{1.15} \setlength{\tabcolsep}{5pt}
\begin{tabular}{l l r r r r r}
\hline
Set & Mmenonic & Total & Ref100 & Recal25 & Ref25 & Early-type \\
\hline
\C\ & Re\textbf{C}al25 & 282 & 154 & 128 &  -- & 46 \\
\F\ & Re\textbf{F}25 & 282 & 154 & -- &  128  & 46 \\
\hline
\end{tabular}}
\end{table}


\subsubsection{Selecting EAGLE galaxies}
\label{SelectingEAGLE.sec}

To enable a proper comparison between our mock observations and the HRS data, we construct a random sample of 282 present-day EAGLE galaxies mimicking the selection criteria described for HRS in Sect.~\ref{HRS.sec}. \ch{Because we have not yet produced any mock observations from which to derive galaxy properties, we use} the intrinsic galaxy properties provided in the public EAGLE database \citep{McAlpine2016}. We first restrict our sample to galaxies with a minimum stellar mass of $10^{9.4}\,\Msun$ for galaxies drawn from the Ref100 snapshot, and $10^{8.5}\,\Msun$ for galaxies drawn from the Recal25 and Ref25 snapshots. These mass cutoffs ensure a minimum numerical resolution of roughly 2000 stellar particles per galaxy, as can be seen from the initial particle masses in Table~\ref{eaglesims.tab}, taking into account the mass transfer due to feedback processes over the lifetime of the stellar populations represented by the particles. As a consequence, our selection favours the high-resolution snapshots Recal25 or Ref25 for galaxies at the lower end of the mass range.

We then use the galaxy-type-dependent $K$-band selection criteria described in Sect.~\ref{HRS.sec}, assuming that all EAGLE galaxies are placed at a distance of 20~Mpc (the median distance of the HRS sample). We employ the specific star-formation rate $\dot{M}_*/M_*$ (sSFR) as a simple proxy for galaxy type, considering galaxies with $\mathrm{sSFR}<10^{-11}~\mathrm{yr}^{-1}$ to be early-type \citep[see, e.g., Fig.~8 in the review by][]{Kennicutt2012}. Finally, we randomly reject galaxies until the sample matches the $K$-band luminosity distribution of the HRS sub-sample studied by \citet{Cortese2012}, as shown in Fig.~\ref{KBandHistogram.fig}.

In fact, we construct two sets of EAGLE galaxies, named \C\ and \F, that each match these criteria. Table~\ref{GalaxySets.tab} and Fig.~\ref{KBandHistogram.fig} illustrate the make-up of these sets. Both sets contain \emph{the same} collection of 154 galaxies drawn from the Ref100 snapshot, including 46 early-type galaxies. In addition, set \C\ includes 128 galaxies drawn from the Recal25 snapshot, and set \F\ likewise includes 128 galaxies drawn from the Ref25 snapshot. These two additional subsets have an identical $K$-band luminosity distribution, and contain no early-type galaxies. \ch{For the lower luminosity bins, the mix of galaxies drawn from the various snapshots is limited by the lack of low-mass galaxies in the Ref100 snapshot, as indicated earlier. For the higher luminosity bins, the mix is limited by the number of galaxies available for each bin in the smaller-volume Recal25 or Ref25 snapshots.}

Our analysis in Sect.~\ref{Results.sec} is mostly based on set \C. However, we evaluate the effects of the recalibration and numerical resolution of the EAGLE simulations by also investigating some of the key results for set \F.


\begin{figure}
  \includegraphics[width=\columnwidth]{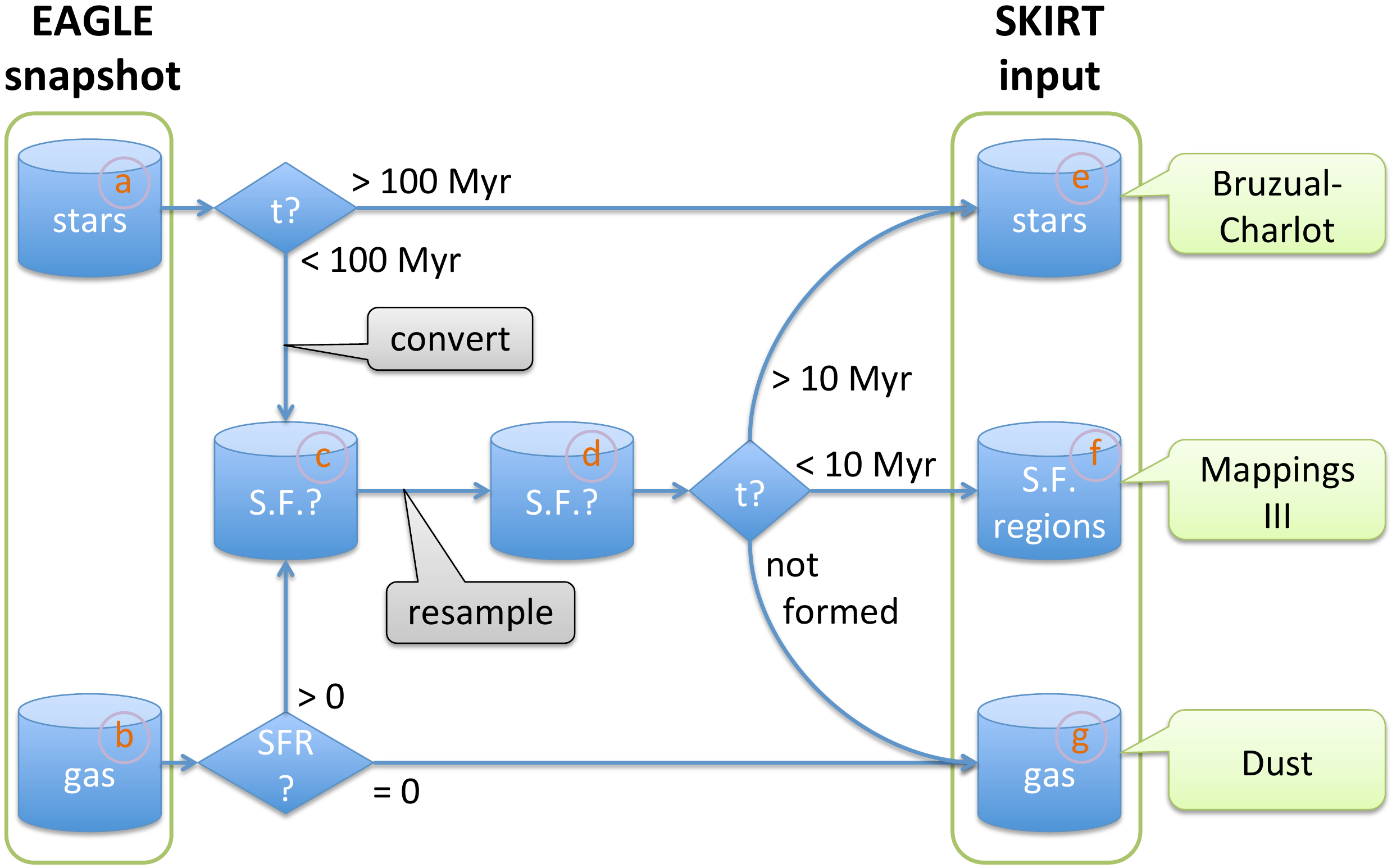}
  \caption{Schematic overview of the procedure used for preparing EAGLE galaxies for \SKIRT. See the text in Sect.~\ref{Preparing.sec} for more details.}
  \label{PreparingGalaxies.fig}
\end{figure}


\subsection{The radiative transfer code \SKIRT}
\label{SKIRT.sec}

\SKIRT\footnote{\SKIRT\ documentation: http://www.skirt.ugent.be\\ \SKIRT\ code repository: https://github.com/skirt/skirt} \citep{Baes2011, Camps2015a} is a public multi-purpose 3D Monte Carlo dust radiative transfer code for simulating the effect of dust on radiation in astrophysical systems. It offers full treatment of absorption and multiple anisotropic scattering by the dust, computes the temperature distribution of the dust and the thermal dust re-emission self-consistently, and supports stochastic heating of dust grains \citep{Camps2015b}. The code handles multiple dust mixtures and arbitrary 3D geometries for radiation sources and dust populations, including grid- or particle-based representations generated by hydrodynamical simulations. The dust density distribution is discretised using one of the built-in dust grids, including state-of-the art octree \citep{Saftly2013}, $k$d-tree \citep{Saftly2014} and Voronoi \citep{Camps2013} grids. The wide range of built-in components can be configured to construct complex models in a parameter file or through a user-friendly interface \citep{Camps2015a, Baes2015}.

While \SKIRT\ is predominantly used to study dusty galaxies \citep[e.g.,][]{Baes2002, Baes2010, Baes2011, DeLooze2012, DeGeyter2014,Saftly2015}, it has also been applied to active galactic nuclei \citep{Stalevski2012}, molecular clouds \citep{Hendrix2015}, and stellar systems \citep{Deschamps2015}. In this work, we use the \SKIRT\ code to perform dust radiative transfer simulations on selected EAGLE galaxies, producing integral field spectroscopy (IFS) data cubes and spectral energy distributions (SEDs) from UV to submm wavelengths as further described in the following sections.

\subsection{Preparing EAGLE galaxies for \SKIRT}
\label{Preparing.sec}

\subsubsection{Extracting galaxies from an EAGLE snapshot}
\label{Extracting.sec}

\citet{Trayford2016} present a procedure for modelling EAGLE galaxies from optical to near-infrared wavelengths ($0.28-2.5~\micron$) using \SKIRT, generating spectra, broad-band photometry, line indices, and multi-band images for a large population of galaxies at redshift $z = 0.1$. We follow the same procedure, paying attention to dust emission and producing spectra and photometry over a much broader wavelength range ($0.02-2000~\micron$).

Figure~\ref{PreparingGalaxies.fig} illustrates the overall process of extracting the data for a galaxy from the EAGLE snapshot and preparing them for post-processing in \SKIRT. For our purposes, a galaxy in an EAGLE snapshot is defined as a gravitationally bound substructure in a halo of dark and baryonic matter represented by particles. These structures are identified by the friends-of-friends and {\large\scshape subfind} \citep{Springel2001,Dolag2009b} algorithms, which are run on the output of the EAGLE simulations. To study a particular galaxy, we extract the corresponding sets of star particles and gas particles (items \emph{a} and \emph{b} in Fig.~\ref{PreparingGalaxies.fig}). Following the convention used by \citet{Schaye2015}, any particles outside a spherical aperture with radius of 30~kpc are ignored. The origin of the coordinate system is positioned at the galaxy's stellar centre of mass. Unless noted otherwise, we retain the galaxy's original orientation, resulting in a `random' viewing angle. In those few cases where we study the results for specific viewing angles, the face-on view looks down from the positive net stellar angular momentum vector of the galaxy, and the edge-on view observes from an arbitrary direction perpendicular to this vector.


\begin{table}
\caption{Input parameters of the SKIRT radiative transfer model for each type of EAGLE particle, in addition to the particle position. The procedure for deriving a dust distribution from the gas particles (item \emph{g} in Fig.~\ref{PreparingGalaxies.fig}) is discussed in Sect.~\ref{DustFraction.sec}. The procedures for the particles representing stellar populations and star-forming regions (items \emph{e} and \emph{f} in Fig.~\ref{PreparingGalaxies.fig}) are discussed in Sect.~\ref{SEDs.sec}.}
\label{Parameters.tab}
{\renewcommand{\arraystretch}{1.15} \setlength{\tabcolsep}{3.5pt}
\begin{tabular}{l l l}
\hline
Param & Description & Origin \\
\hline
&\kern-2em\emph{Dust distribution} \\
$h$ & Smoothing length & Particle \\
$M$ & Current gas mass & Particle \\
$Z$ & Gas metallicity & Particle \\
$T$ & Temperature of the gas & Particle \\
SFR & Star-formation rate of the gas & Particle \\
$T_\mathrm{max}$ & Highest temperature at which gas contains dust & Preset value \\
$f_\mathrm{dust}$ & Fraction of the metallic gas locked up in dust & Free param \\[1mm]
&\kern-2em\emph{Young and evolved stars} \\
$h$ & Smoothing length & Particle \\
$M_\mathrm{init}$ & Birth mass of the stellar population & Particle \\
$Z$ & Metallicity of the stellar population & Particle \\
$t$ & Age of the stellar population  & Particle\\[1mm]
&\kern-2em\emph{Star-forming regions} \\
$h$ & Smoothing length & Calculated \\
$M$ & Mass of the \HII\ region & Sampled \\
SFR & Star-formation rate of the \HII\ region & Calculated \\
$Z$ & Metallicity of the \HII\ region & Parent particle \\
$\rho$ & Gas density at the \HII\ region's position & Parent particle \\
$P$ & Pressure of the ambient ISM & Calculated \\
$C$ & Compactness of the \HII\ region & Calculated \\
$f_\mathrm{PDR}$ & Dust covering fraction of the PDR region & Free param \\
\hline
\end{tabular}}
\end{table}


\subsubsection{Re-sampling star-forming regions}
\label{Resampling.sec}

Star formation in EAGLE occurs stochastically: at each time step, a gas particle has a certain probability of being wholly converted to a star particle. Because individual particle masses are rather high (of the order of $10^6~\Msun$ for the reference simulation, see Table~\ref{eaglesims.tab}), a typical EAGLE galaxy contains only a small number of young star particles, unrealistically clumping all of the galaxy's young stars in a few point-like regions. This introduces sampling issues, which we alleviate by reprocessing the star-forming gas particles and the youngest star particles before feeding them into the \SKIRT\ radiative transfer code, as illustrated in Fig.~\ref{PreparingGalaxies.fig}.

As a first step, we build a set of star-forming region candidates (item \emph{c} in Fig.~\ref{PreparingGalaxies.fig}), including all star particles younger than 100~Myr \ch{-- the typical timescale of a starburst \citep{Groves2008} --} and all gas particles with a nonzero star-formation rate (SFR). All other particles, i.e.\ older star particles and non-star-forming gas particles, are transferred directly to the corresponding \SKIRT\ input sets (items \emph{e} and \emph{g} in Fig.~\ref{PreparingGalaxies.fig}). The young star particles are converted back to star-forming gas particles. The SFR at the time of birth of these particles is calculated using the relation between pressure and SFR described in Sect.\ 4.3 of \citet{Schaye2015} and originally in \citet{Schaye2008}, which is based on the empirical Kennicutt-Schmidt law \citep{Kennicutt1998}.

In the second step, the star-forming region candidates are re-sampled into a number of sub-particles (item \emph{d} in Fig.~\ref{PreparingGalaxies.fig}) with lower masses drawn randomly from the power-law mass distribution function,
\begin{equation}
\frac{\mathrm{d}N}{\mathrm{d}M}\propto M^{-1.8}\quad \mathrm{with}~M\in[700,10^6]\,\Msun.
\label{mass-distrib.eq}
\end{equation}
This distribution of masses is inspired by observations of molecular clouds in the Milky Way reported by \citet{Heyer2001} and reviewed in Sect.~2.5 of \citet{Kennicutt2012}. Once a sufficient number of sub-particles have been generated to approximately represent the parent particle's mass, the sub-particle masses are proportionally adjusted to ensure exact mass conservation. The resulting sub-particles are assigned a formation time sampled randomly to represent their parent's SFR and mass, assuming a constant SFR over a 100~Myr lifetime. The sub-particles that formed less than 10~Myr ago \ch{-- the typical lifetime of an \HII\ region \citep{Groves2008} --} are placed into a new \SKIRT\ input set defining star-forming regions (item \emph{f}); those that formed more than 10~Myr ago are recast as star particles (item \emph{e}); and those that have not yet formed are recast as gas particles (item \emph{g}).

Finally, the smoothing lengths and positions of the star-forming sub-particles are adjusted to match our post-processing assumptions as explained in Sect.~\ref{SEDs.sec}.


\begin{figure}
  \includegraphics[width=0.99\columnwidth]{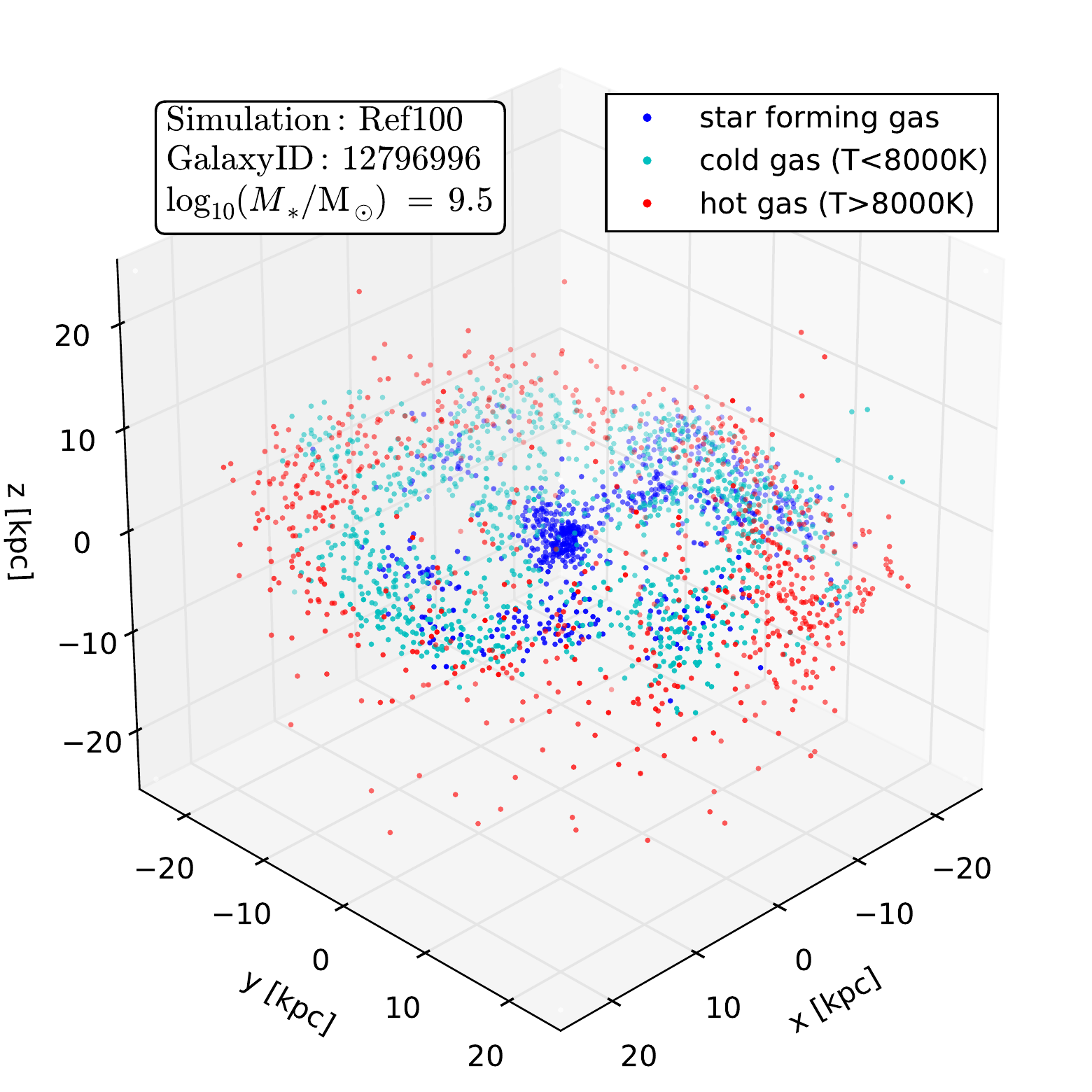}
  \caption{A projection of the gas particle positions in an EAGLE disc galaxy hand-picked for illustrative purposes. Our procedure allocates dust for star-forming gas (blue) and for cold gas (cyan). Hot gas (red) is deemed not to contain any dust.}
  \label{GasParticles.fig}
\end{figure}


\subsubsection{Deriving the diffuse dust distribution}
\label{DustFraction.sec}

Table~\ref{Parameters.tab} offers an overview of the parameters defining the SKIRT radiative transfer model for each type of input particles, as discussed in the current and the following section. We derive a dust mass, $M_\mathrm{dust}$, for each particle in \SKIRT's `gas' input set (item \emph{g} in Fig.~\ref{PreparingGalaxies.fig}) according to
\begin{equation}
M_\mathrm{dust} = \begin{cases}
f_\mathrm{dust}\,Z\,M & \mathrm{if}\; T<T_\mathrm{max} \;\mathrm{or}\; \mathrm{SFR}>0  \\
0 & \mathrm{otherwise,}
\end{cases}
\label{dustfromgas.eq}
\end{equation}
where $Z$, $M$, $T$, and SFR are the metallicity (metal mass fraction)\footnote{\label{metallicity.fn}We use the SPH smoothed metallicity rather than the particle metallicity; see \citet{Wiersma2009b} and \citet{Schaye2015} for more information.}, current mass, temperature, and star-formation rate given by the gas particle's properties in the EAGLE snapshot, and $f_\mathrm{dust}$ and $T_\mathrm{max}$ are free parameters. The characterisation of gas particles based on the conditions of Eq.~(\ref{dustfromgas.eq}) is illustrated in Fig.~\ref{GasParticles.fig} for an EAGLE disc galaxy. The star-forming (blue) and cold (cyan) gas particles trace the spiral arms in the galactic disk, while the hot gas (red) is located in the outskirts, as expected.

In summary, Eq.~(\ref{dustfromgas.eq}) assumes that a constant fraction $f_\mathrm{dust}$ of the metallic gas is locked up in dust, as long as the gas is forming stars or the gas is colder than the cutoff temperature $T_\mathrm{max}$. The assumption of a fixed dust-to-metal fraction $f_\mathrm{dust}$ is observed to be an appropriate approximation for a variety of environments \citep{Dwek1998, James2002, Brinchmann2013, Zafar2013}. We will vary this parameter as part of our analysis in Sect.~\ref{Results.sec}.

The condition $\mathrm{SFR}>0$ captures the re-sampled gas particles that are eligible for star formation but were not actually converted into a star-forming region (the `not formed' arrow between items \emph{d} and \emph{g} in Fig.~\ref{PreparingGalaxies.fig}). We need this condition because the EAGLE simulations assign a non-physical temperature to star-forming gas particles (see Sect.~\ref{EAGLE.sec}). However, by definition, the star-forming gas can be assumed to be sufficiently cold to form dust. 

The temperature cutoff \,$T<T_\mathrm{max}$ for the non-star-forming gas particles accounts for the fact that dust cannot form, or is rapidly destroyed, in hot gas \citep[e.g.,][]{Guhathakurta1989}. We need to determine an appropriate temperature cutoff value. Unfortunately, since the EAGLE simulations do not model the cold gas phase in the ISM (see Sect.~\ref{EAGLE.sec}), we cannot properly constrain $T_\mathrm{max}$ using a physically motivated procedure. For our analysis in Sect.~\ref{Results.sec}, we select a value of $T_\mathrm{max}=8000~\mathrm{K}$, corresponding to the value of $T_\mathrm{eos}$ at $n_\mathrm{H}=0.1~\mathrm{cm}^{-1}$ used in the EAGLE simulations \citep[Sect.~4.3 of ][]{Schaye2015}.


\begin{figure}
  \includegraphics[width=0.995\columnwidth]{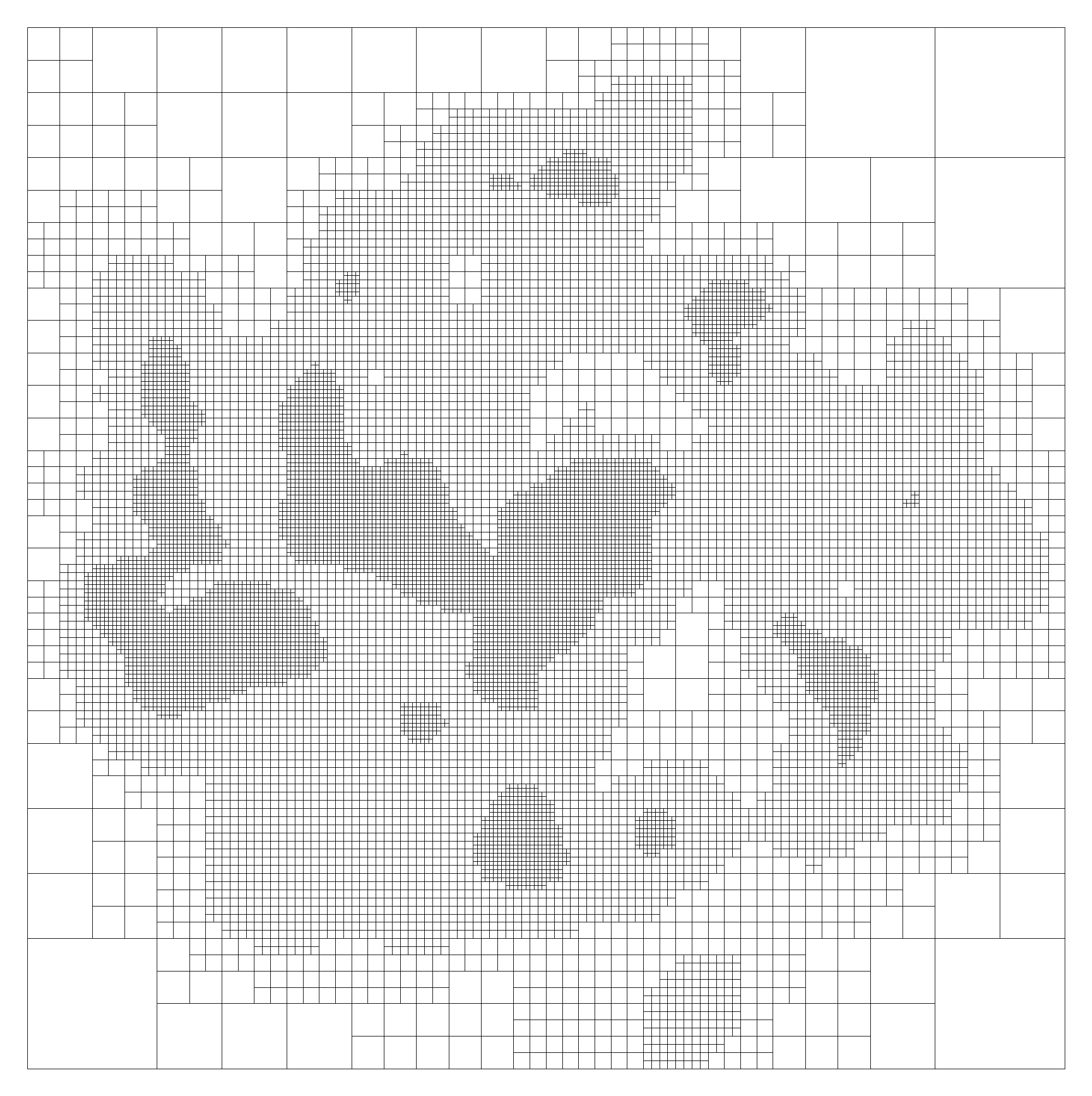}
  \caption{A cut along the galactic plane through an octree dust grid constructed for the EAGLE galaxy shown in Fig.~\ref{GasParticles.fig}. The darker areas trace regions of higher dust density (the grid has smaller dust cells and thus more cell boundaries). For presentation purposes, the illustrated grid uses fewer refinement levels and covers a smaller aperture than the grid actually used by \SKIRT\ in this work \ch{(25 kpc radius rather than 30 kpc)}.}
  \label{DustGrid.fig}
\end{figure}

\begin{figure*}
  \includegraphics[width=\textwidth]{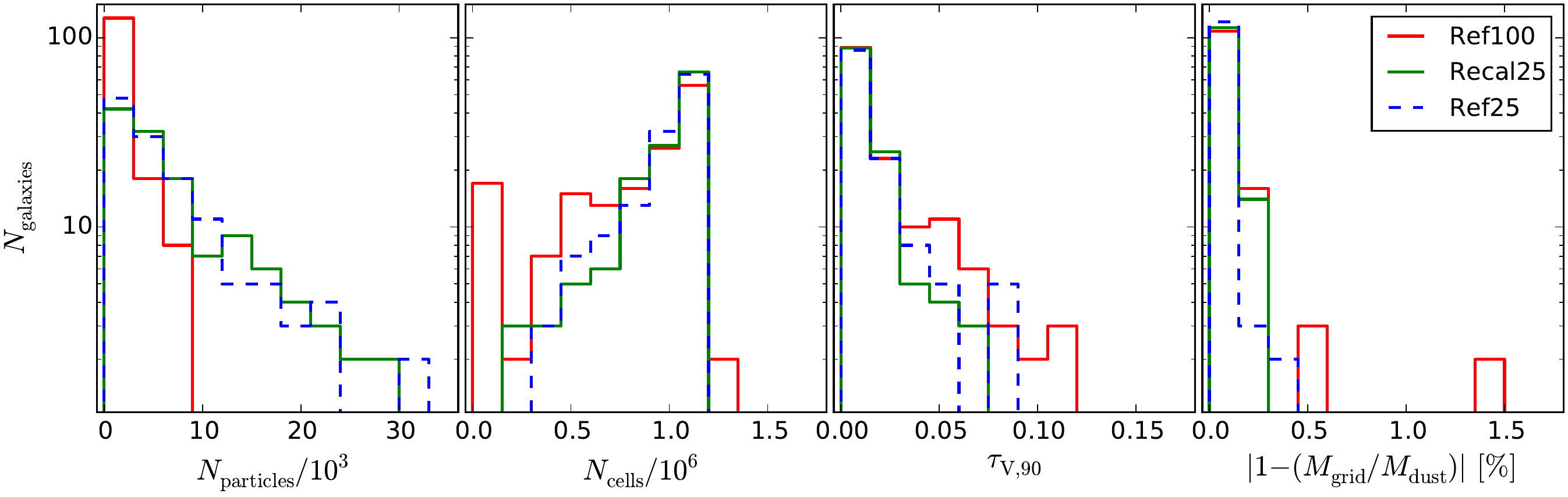}
  \caption{Distribution of \SKIRT\ dust discretisation properties for the EAGLE galaxies analysed in this work (see Table~\ref{GalaxySets.tab}), processed with $f_\mathrm{dust}=0.3$. From left to right: the number of gas particles that include dust (i.e.\ cold or star-forming gas particles); the number of cells in the dust grid constructed by \SKIRT; the 90\% percentile V-band optical depth of the cells in the dust grid; the discretisation error on the total dust mass (i.e.\ difference between the dust mass in the grid and in the incoming particles).}
  \label{DustGridProperties.fig}
\end{figure*}


\subsubsection{Assigning SEDs to particles}
\label{SEDs.sec}

Each particle in \SKIRT's `stars' input set (item \emph{e} in Fig.~\ref{PreparingGalaxies.fig}) is assigned a stellar population SED from the \citet{Bruzual2003} family, using the birth mass, metallicity$^{\ref{metallicity.fn}}$ and age given by the particle properties in the EAGLE snapshot (see Table~\ref{Parameters.tab} for an overview). We use the low resolution version of the Padova1994/Chabrier model, which is one of the two models recommended by \citet{Bruzual2003}.

For the particles in the `star-forming regions' input set (item \emph{f} in Fig.~\ref{PreparingGalaxies.fig}) we follow the procedure described by \citet{Jonsson2010}. Each particle is assigned an appropriate starburst SED from the MAPPINGS III family \citep{Groves2008}. These templates model both the \HII\ region and the photodissociation region (PDR) surrounding the star-forming core, including the dust contained in those regions. We attempt to compensate for the additional dust mass assumed by the MAPPING III model by removing the equivalent amount of dust from the diffuse dust component. We now discuss this process in more detail.

The MAPPINGS III templates are parametrised by the SFR and the metallicity of the star-forming region, the pressure of the ambient ISM, the \HII\ region compactness, and the covering fraction of the associated PDR (see Table~\ref{Parameters.tab} for an overview). The SFR is determined from the mass assigned to the star-forming particle (as discussed in Sect.~\ref{DustFraction.sec}), assuming a constant SFR during the \HII\ region's lifetime of 10~Myr \citep[following][]{Groves2008}. The metallicity, $Z$, is taken directly from the particle properties$^{\ref{metallicity.fn}}$ in the EAGLE snapshot. The ambient pressure of the ISM, $P$, is calculated from the particle's density, $\rho$, using the polytropic equation of state imposed on star-forming particles (see Sect.~\ref{EAGLE.sec}). The \HII\ region compactness, $C$, is designed to reflect the dust temperature distribution in the \HII\ region (time-averaged over its lifetime), so that it predominantly controls the form of the FIR continuum dust emission. In our procedure the value of this parameter is derived from the ambient pressure, $P$, and our assigned particle mass, $M$ (see Eq.~\ref{mass-distrib.eq}), using Eq.~(13) of \citet{Groves2008}, i.e.,
\begin{equation}
\log_{10} C = \frac{3}{5}\log_{10}\left(\frac{M}{\Msun}\right) + \frac{2}{5}\log_{10}\left(\frac{P/k_\mathrm{B}}{\mathrm{cm}^{-3}\,\mathrm{K}}\right),
\label{compactness.eq}
\end{equation}
where $k_\mathrm{B}$ is the Boltzmann constant. Finally, the parameter $f_\mathrm{PDR}$ is defined as the time-averaged dust covering fraction of the photodissociation region (PDR) surrounding the star-forming core over the \HII\ region's lifetime. Starbursts in which the PDR's dust entirely envelops the \HII\ region have $f_\mathrm{PDR}=1$, while uncovered \HII\ region complexes have $f_\mathrm{PDR}=0$. The covering fraction is treated as a free parameter, which we will vary as part of our analysis in Sect.~\ref{Results.sec}. 

\ch{As indicated above, the MAPPINGS III templates model the dust residing in the PDR region in addition to the core \HII\ region itself.} Following \citet{Jonsson2010}, we consider the region represented by the MAPPINGS III templates (including PDR and \HII\ region) to be ten times as massive as the star-forming core represented by the particle. To determine the spatial extent of the region's emission, we assume that the region's centre has the same density as the local ambient ISM. For the cubic spline kernel employed in SKIRT, this leads to the easily inverted relation, $10M = (\pi/8)\rho h^3$, between the \HII\ region mass, $M$, the ambient density, $\rho$, and the particle smoothing length, $h$. We also randomly shift the positions of the star-forming sub-particles within the smoothing sphere of the parent particle (see Sect.~\ref{Resampling.sec}) to avoid overlap between the modelled regions.

\ch{To avoid double counting the dust residing in the PDR region,} we subtract this PDR dust from the diffuse dust distribution derived as discussed in Sect.~\ref{DustFraction.sec}. We insert a `ghost' gas particle with negative mass in the \SKIRT\ gas input set (see Sect.~\ref{DustFraction.sec}) for each star-forming particle. The ghost particle receives the (negative) mass of the corresponding PDR region, i.e.\ ten times the mass of the star-forming particle. When sampling the gas (or dust) density field, \SKIRT\ combines the negative ghost densities with the positive densities defined by the other particles, clipping the total density to zero if needed. To lower the probability of this occurring, we artificially increase the smoothing length of the ghost particle by a factor of three. According to our tests, this sufficiently alleviates the issue without otherwise affecting the results.

\subsection{Radiative transfer on EAGLE galaxies}
\label{RT.sec}

This section describes the \SKIRT\ configuration used to perform the radiative transfer simulations on the EAGLE galaxies.

\subsubsection{Dust grid}
\label{DustGrid.sec}

The \SKIRT\ radiative transfer procedure requires the dust density distribution of the system under study to be discretised over a dust grid. Within each grid cell, the dust density and all other physical quantities, such as the radiation field, are assumed to be constant. \SKIRT\ implements a performance-optimised mechanism to calculate the dust mass in each grid cell from the smoothed particles defining a galaxy. The particles are interpolated using a scaled and truncated Gaussian kernel designed to approximate a finite-support cubic spline kernel \citep{Altay2013,Baes2015}.

Here, we use an adaptive, hierarchical Cartesian grid that encloses the 30~kpc aperture considered for each galaxy (see Sect.~\ref{Extracting.sec}). Specifically, we use an octree grid \citep{Saftly2013} that automatically subdivides cells until each cell contains less than a fraction $\delta_\mathrm{max}=3\times10^{-6}$ of the total dust mass in the model, with a maximum of 10 subdivision levels (see Fig.~\ref{DustGrid.fig}). The smallest possible cell is thus about 60 pc on a side, which offers 5-10 times better resolution than the typical gravitational softening length in the EAGLE simulations (see Table~\ref{eaglesims.tab}).

Figure~\ref{DustGridProperties.fig} provides some relevant statistics on the discretisation of the diffuse dust density for the EAGLE galaxies analysed in this work. The leftmost panel shows that about half of the galaxies in our selection have more than 3000 gas particles that include dust, i.e.\ particles representing cold or star-forming gas (see Sect.~\ref{DustFraction.sec}), which is sufficient to spatially resolve the diffuse dust distribution. Further analysis (not shown) indicates that about 100 of our galaxies have less than 100 `dusty' gas particles, however this includes most of the early-type galaxies, which do not contain much dust anyway.

The two middle panels of Figure~\ref{DustGridProperties.fig} show properties of the dust grids constructed by \SKIRT\ to perform radiative transfer on our EAGLE galaxies. Most dust grids have more than 250\,000 cells, which is more than sufficient to resolve the imported smoothed particles. Also, over 90\% of the dust cells in each grid have a V-band optical depth of less than 0.12 (and most have much lower optical depth), indicating that the grid properly resolves even the densest regions in the dust mass.

The rightmost panel of Figure~\ref{DustGridProperties.fig} shows the difference between the dust mass obtained by summing over all cells in the dust grid, and the dust mass obtained by summing over the incoming particles. For most galaxies, this dust discretisation error is limited to less than a third of a per cent, with some outliers of up to 1.5 per cent. While part of this error is caused by grid resolution limitations, further analysis (not shown) indicates that the larger discrepancies in the outliers are caused by the negative dust masses which are introduced to compensate for the dust modeled by star-forming regions (see Sect.~\ref{SEDs.sec}). Specifically, the imported dust density becomes negative in some areas, and is then clipped to zero when building the dust grid.


\begin{figure*}
  \includegraphics[width=\textwidth]{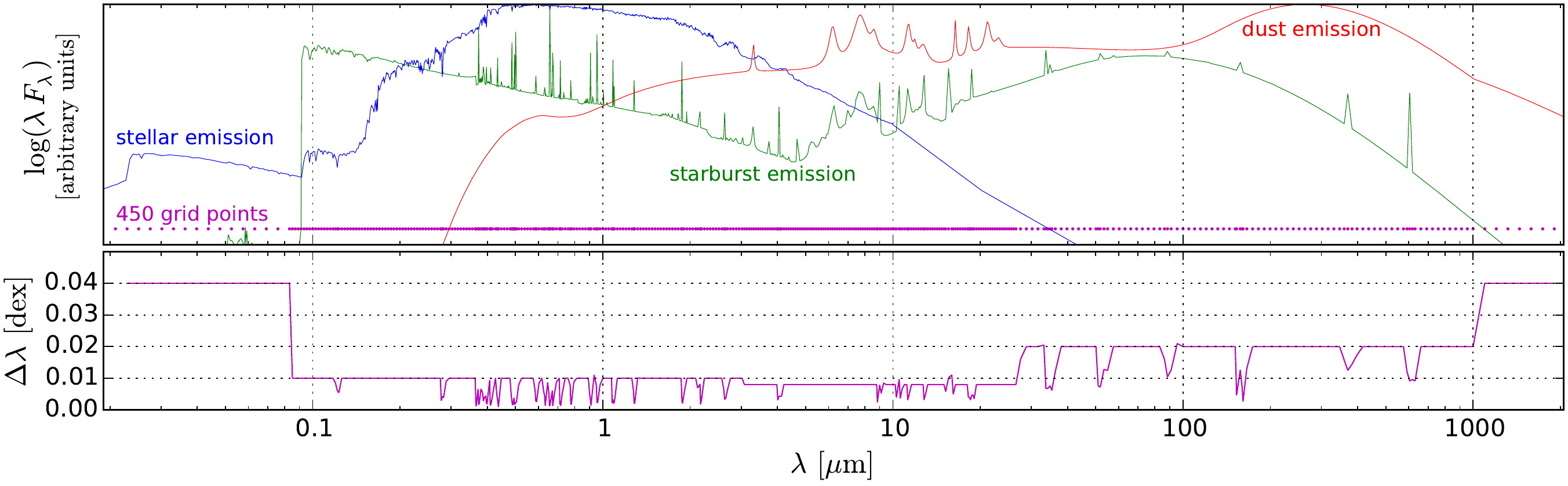}
  \caption{Characteristics of the wavelength grid used in all \SKIRT\ calculations for this work. The three curves in the top panel illustrate typical SEDs for an evolved stellar population (blue), a star-forming region including young stellar objects and dust (green), and stochastically heated diffuse dust (red), plotted on an arbitrary logarithmic scale. The dots (magenta) represent the wavelength grid points. The curve in the bottom panel (magenta) indicates the distance between successive wavelength points on a logarithmic scale.}
  \label{WavelengthGrid.fig}
\end{figure*}


\subsubsection{Dust model}
\label{DustModel.sec}

To represent the \ch{diffuse} dust in the EAGLE galaxies, we use \ch{the model presented by \citet{Zubko2004}. This model includes} a dust mixture of non-composite graphite and silicate grains and neutral and ionised polycyclic aromatic hydrocarbon (PAH) molecules, designed so that the global dust properties accurately reproduce the extinction, emission and abundance constraints of the Milky Way. The optical properties are taken from Bruce Draine's website\footnote{\url{http://www.astro.princeton.edu/~draine/dust/dust.diel.html}} \citep{Draine1984, Laor1993, Li2001}. The calorimetric properties follow the prescription of \citet{Draine2001}. The grain size distributions for each population are taken from \citet{Zubko2004}.

The dust emission spectrum is calculated for each dust cell based on the stellar radiation absorbed by the dust in that cell. The calculation includes the effects of stochastically heated grains, i.e.\ dust grains and PAH molecules that are not in local thermal equilibrium with the radiation field, using the scheme described by \citet{Camps2015b}. To facilitate this calculation, \SKIRT\ discretises the size range of the dust grains into several size bins, for each type of grain material separately. For this work, following the recommendations of \citet{Camps2015b}, \SKIRT\ uses 15 size bins for each of the graphite and silicate components, and 10 size bins for each of the neutral and ionised PAHs.

\ch{The MAPPINGS III templates, used in our post-processing procedure to model star-forming regions (Sect.~\ref{SEDs.sec}), employ a dust model with optical properties taken from the same sources \citep{Laor1993, Li2001}. The grain size distribution and the treatment of PAHs \citep{Groves2008, Dopita2005} are similar but not identical to our model for the diffuse dust. While we don't believe these differences substantially affect our analysis and conclusions, it is important to note the uncertainties involved with the selection of a dust model.}

\subsubsection{Wavelength discretisation}
\label{WavelengthGrid.sec}


\begin{table}
\caption{Evaluation of the wavelength grid and numerical convergence for the \SKIRT\ simulations in this work. The first two columns list the name of the instrument for which mock broad-band fluxes are calculated and the corresponding pivot wavelength according to Eqs.~(\ref{pivot.eq}) or (\ref{pivotbol.eq}). The next three columns show the differences between the magnitude calculated on a high-resolution wavelength grid and on our default wavelength grid, for the three SEDs shown in Fig.~\ref{WavelengthGrid.fig}. The last column shows the maximum magnitude differences for SEDs calculated from \SKIRT\ simulations with different dust grid resolutions and numbers of photons. The dashed line separates photon counters (top) and bolometers (bottom).}
\label{GridMagnitudeErrors.tab}
{\renewcommand{\arraystretch}{1.05} \setlength{\tabcolsep}{4.5pt}
\begin{tabular}{l r r r r r}
\hline
 &  & \multicolumn{3}{c}{-------- Wavelength grid --------} & Dust grid \\
Band & $\lambda_\mathrm{pivot}$ & SF region & Stellar & Dust & \& photons \\
     & (\micron) & ($\Delta$\,mag) & ($\Delta$\,mag) & ($\Delta$\,mag) & ($\Delta$\,mag) \\
\hline
GALEX FUV  & 0.1535 & 0.002 & 0.041 &    -- & 0.004 \\
GALEX NUV  & 0.2301 & 0.003 & 0.007 &    -- & 0.003 \\
SDSS u     & 0.3557 & 0.047 & 0.027 & 0.001 & 0.002 \\
SDSS g     & 0.4702 & 0.087 & 0.005 & 0.001 & 0.001 \\
SDSS r     & 0.6176 & 0.054 & 0.003 & 0.001 & 0.002 \\
SDSS i     & 0.7490 & 0.011 & 0.002 & 0.002 & 0.002 \\
SDSS z     & 0.8947 & 0.016 & 0.005 & 0.001 & 0.001 \\
2MASS $J$  & 1.239  & 0.016 & 0.014 & 0.011 & 0.001 \\
2MASS $H$  & 1.649  & 0.011 & 0.012 & 0.003 & 0.002 \\
2MASS $K_S$ & 2.164 & 0.003 & 0.013 & 0.002 & 0.001 \\
WISE W1    & 3.390  & 0.021 & 0.003 & 0.018 & 0.001 \\
WISE W2    & 4.641  & 0.005 & 0.005 & 0.001 & 0.001 \\
WISE W3    & 12.57  & 0.001 & 0.001 & 0.001 & 0.003 \\
WISE W4    & 22.31  & 0.001 & 0.001 & 0.009 & 0.003 \\
\hdashline
MIPS 24 & 23.59 & 0.001 & 0.001 & 0.008 & 0.003 \\
MIPS 70 & 70.89 & 0.001 & 0.001 & 0.001 & 0.003 \\
MIPS 160 & 155.4 & 0.001 & 0.001 & 0.001 & 0.004 \\
PACS 70    & 70.77  & 0.001 & 0.001 & 0.001 & 0.002 \\
PACS 100   & 100.8  & 0.001 & 0.001 & 0.001 & 0.003 \\
PACS 160   & 161.9  & 0.001 & 0.008 & 0.001 & 0.004 \\
SPIRE 250 ext & 252.5  & 0.001 &    -- & 0.001 & 0.034 \\
SPIRE 350 ext & 354.3  & 0.001 &    -- & 0.001 & 0.034 \\
SPIRE 500 ext & 515.4  & 0.026 &    -- & 0.001 & 0.035 \\
\hline
\end{tabular}}
\end{table}



\begin{figure}
  \includegraphics[width=\columnwidth]{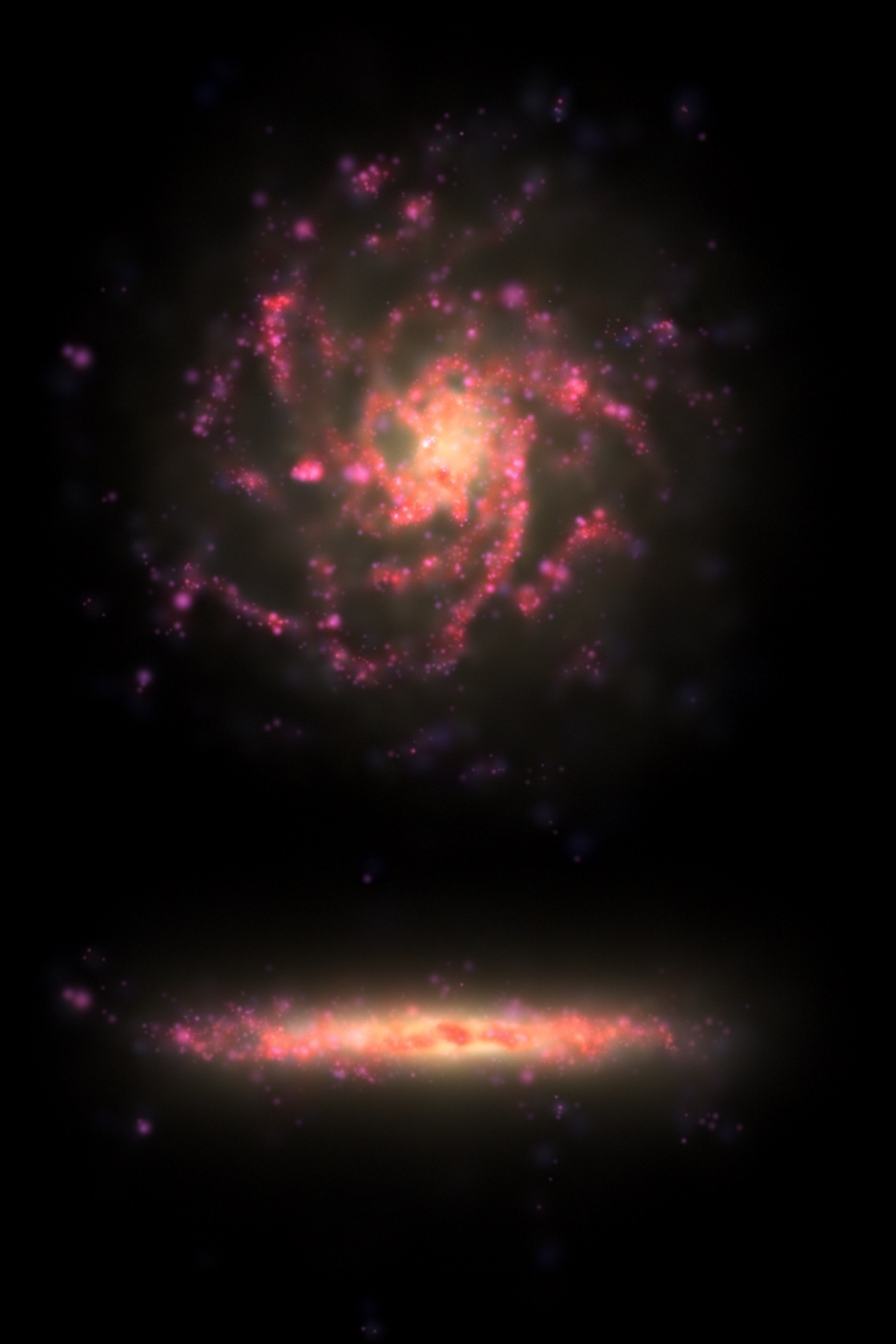}
  \caption{\ch{Face-on (top) and edge-on (bottom) views of a Milky Way-like EAGLE galaxy (Ref25; GalaxyID 639646; $M_*=1.75\times10^{10}~\Msun$) post-processed using our procedures. The images combine optical colours (blue, green and red for SDSS $g$, $r$ and $i$ fluxes) with additional blue for GALEX NUV flux and red for \emph{Herschel} PACS 100~\micron\ flux. The views cover an aperture with 30~kpc radius and their orientation is defined by stellar angular momentum, as described in Sect.~\ref{Extracting.sec}.}}
  \label{GalaxyImages.fig}
\end{figure}

\begin{table}
\caption{Properties of the \emph{Herschel} SPIRE 250/350/500 instruments used in our mock flux derivation. The beam FWHM and beam area are taken from \citet{Ciesla2012}. For the flux limit, we use the confusion noise level from \citet{Nguyen2010}.
}
\label{HerschelProps.tab}
{\renewcommand{\arraystretch}{1.15} \setlength{\tabcolsep}{4.5pt}
\begin{tabular}{l l r r r}
\hline
  & Units &  $250\,\micron$ &  $350\,\micron$  &  $500\,\micron$  \\
\hline
Beam FWHM & arcsec & 18.2 & 24.5 & 36.0 \\
Beam area & arcsec$^2$ & 423 & 751 & 1587 \\
Flux limit & mJy/beam & 5.8 & 6.3 & 6.8 \\
\hline
\end{tabular}}
\end{table}


The \SKIRT\ code employs a single wavelength grid for all calculations. The input SEDs and dust properties are sampled on this grid, photon packages are given wavelengths corresponding to the grid points, dust absorption and re-emission are calculated for the wavelength bins defined by the grid, and the output fluxes are recorded on the same grid.

The wavelength grid used in all \SKIRT\ calculations for this work is illustrated in Fig.~\ref{WavelengthGrid.fig}. It resolves the relevant features in the input SEDs (see Sect.~\ref{SEDs.sec}) and in the emission spectrum of the dust population (see Sect.~\ref{DustModel.sec}). The grid has 450 wavelength points from 0.02 to 2000 \micron\ laid out on a logarithmic scale. The bin widths are 0.04 dex in the outer wavelength ranges where fluxes are low, 0.02 dex in the dust emission continuum, 0.01 dex in the optical range, and under 0.01 dex in the PAH emission range and for specific emission or absorption features in the employed input spectra.

To further inspect this discretisation, we compare band-integrated fluxes (see Appendix~\ref{Photometry.sec}) calculated on our default 450-point wavelength grid with those calculated on a high-resolution grid with 20\,000 points. For this purpose, we select a typical SED for a stellar population, one for a star-forming region, and one for stochastically heated dust (see Fig.~\ref{WavelengthGrid.fig}). We calculate the fluxes for these SEDs in a set of bands essentially covering the complete wavelength range, using Eqs.~(\ref{meanflux.eq}) or (\ref{meanfluxbol.eq}). The results calculated on our 450-point wavelength grid are accurate to within 0.1 mag for all bands, and often much better. The results for the bands used in this work are listed in Table~\ref{GridMagnitudeErrors.tab}.

\subsubsection{Photon packages}
\label{Photons.sec}

The \SKIRT\ radiative transfer simulation proceeds in two phases. In the first phase, \SKIRT\ launches photon packages randomly originating at the stars and the star-forming regions, and traces these packages through the dusty medium. The simulation loop accounts for the effects of scattering off dust grains, and keeps track of the radiation absorbed in each dust cell. After this phase completes, the code calculates the emission spectrum of the dust population in each dust cell based on the established radiation field, taking into account the probabilistic thermal emission of small grains and PAH molecules \citep{Camps2015b}. In the second phase, \SKIRT\ launches photon packages originating from the dust distribution, corresponding to the calculated emission spectra, and traces these packages through the dusty medium as well.

For this work, we instruct SKIRT to ignore dust heating by photon packages emitted from the dust, substantially reducing the calculation time. This is justified because the body of dust in a normal galaxy is essentially transparent to infrared radiation. We verified this assumption for our EAGLE sample by comparing the simulation results with and without dust self-heating for the highest dust-mass galaxies. Finally, we configure \SKIRT\ to launch $5\times10^5$ photon packages for each of the $450$ points in the wavelength grid during each of the two phases. Thus the \SKIRT\ simulation for each EAGLE galaxy traces $4.5\times10^8$ photon packages. In Sect.~\ref{NumConvergence.sec} we confirm that this choice is appropriate.

\subsubsection{Mock fluxes}
\label{Instruments.sec}


\begin{figure*}
  \includegraphics[width=\textwidth]{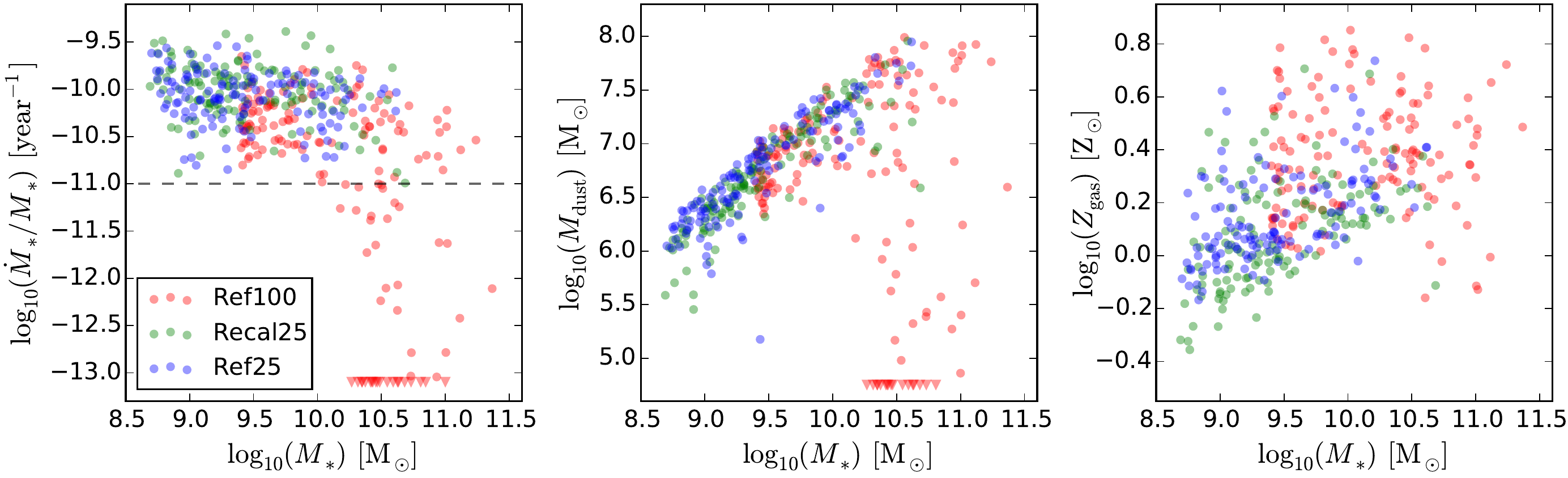}
  \caption{Selected intrinsic properties of the EAGLE galaxies analysed in this work (see Table~\ref{GalaxySets.tab}) plotted versus intrinsic stellar mass. Left panel: specific star-formation rate (sSFR); galaxies \ch{with sSFR below $10^{-11}~\mathrm{yr}^{-1}$ (dashed line)} are deemed to be early-types; galaxies with sSFR below $10^{-13.1}~\mathrm{yr}^{-1}$ are plotted as upper bounds at that value. Middle panel: dust mass, assuming a dust-to-metal fraction $f_\mathrm{dust}=0.3$; galaxies with dust mass below $10^{4.75}\,\Msun$ are plotted as upper bounds at that value. Right panel: overall metallicity of the gas that includes dust, in units of $\Zsun=0.0127$; galaxies with zero dust mass are omitted.}
  \label{IntrinsicProperties.fig}
\end{figure*}


Mock detectors are placed along two of the coordinate axes at a fixed distance of 20 Mpc from the model, using parallel projection. If the model has been properly rotated (see Sect.~\ref{Extracting.sec}), this results in a face-on and an edge-on view of the galaxy. The selected detector distance matches the median distance of the HRS galaxies; see Sect.~\ref{SelectingEAGLE.sec}. Each detector records an integral field data cube (a $400\times400$ pixel frame at each of the wavelength grid points) in addition to the spatially integrated fluxes at each wavelength grid point. From these results, we produce band-integrated fluxes and absolute magnitudes corresponding to the following filters (see also Table~\ref{GridMagnitudeErrors.tab}): GALEX FUV/NUV \citep{Morrissey2007}; SDSS $ugriz$ \citep{Doi2010}; 2MASS $JHK$ \citep{Cohen2003}; WISE W1/W2/W3/W4 \citep{Wright2010}; \emph{Spitzer} MIPS 24/70/160 \citep{Rieke2004}; \emph{Herschel} PACS 70/100/160 \citep{Poglitsch2010}; and \emph{Herschel} SPIRE 250/350/500 for extended sources \citep{Griffin2010}.

To obtain the integrated fluxes, we convolve the simulated SED with the instrument's response curve. The precise procedure depends on whether the instrument counts photons or measures energy (bolometer); the formulae are summarised in Appendix~\ref{Photometry.sec}. The GALEX, SDSS, 2MASS and WISE instruments are photon counters; the \emph{Spitzer} MIPS and the \emph{Herschel} PACS and SPIRE instruments are bolometers.

Because our analysis in Sect.~\ref{Results.sec} relies heavily on the \emph{Herschel} SPIRE 250/350/500 fluxes, and because actual observations in these submm bands suffer from fairly severe observational limitations, we perform an additional procedure for these fluxes. Table~\ref{HerschelProps.tab} lists the relevant instrument properties, taken from \citet{Ciesla2012} and \citet{Nguyen2010}. We first perform a convolution with the corresponding instrument response function for each of the $400\times400$ pixels in the recorded frames. \ch{We then convolve the resulting frame (spatially) with a Gaussian filter scaled to the full width at half maximum (FWHM) of the instrument's beam, and re-bin the pixels in the frame to match the beam area of the instrument.} From this frame, we eliminate all pixels with a flux value below the sensitivity level of the instrument, and we finally sum over the remaining pixels to obtain the total flux.

\ch{While the analysis in this work uses spatially integrated fluxes, it is instructive to examine images illustrating the results of our procedures. Fig.~\ref{GalaxyImages.fig} shows face-on and edge-on views of a Milky Way-like EAGLE galaxy post-processed as described in this methods section. The images combine an optical view using SDSS $g$, $r$ and $i$ fluxes with additional blue for GALEX NUV flux and red for \emph{Herschel} PACS 100~\micron\ flux. To obtain these fluxes, the data cubes recorded by SKIRT were convolved, pixel by pixel, with the response curve for each instrument. The resulting purple colours indicate star-forming regions, which strongly emit both in the NUV and FIR. The red colours indicate bodies of diffuse interstellar dust.}

\subsubsection{Numerical convergence}
\label{NumConvergence.sec}

A numerical convergence test can help ascertain that our discretisation settings are appropriate. To this end, we perform the \SKIRT\ simulations for galaxy set \C\ (Table~\ref{GalaxySets.tab}) using a higher-resolution dust grid and shooting more photon packages than for our default setup. Specifically, we set the maximum mass fraction per cell to $\delta_\mathrm{max}=2\times10^{-6}$ rather than $3\times10^{-6}$ (see Sect.~\ref{DustGrid.sec}), and we increase the number of photon packages launched per wavelength grid point to $10^6$ from $5\times10^5$ (see Sect.~\ref{Photons.sec}). We then calculate the fluxes in the various bands used for this work according to the procedure described in Sect.~\ref{Instruments.sec}, and we compare the results from the high-resolution simulation with those from the default setup. The rightmost column in Table~\ref{GridMagnitudeErrors.tab} shows the absolute value of the resulting magnitude differences. The fluxes are accurate to within 0.05 mag for all bands, and even to within 0.005 mag for all but the four longest-wavelength bands. The somewhat larger errors for the \emph{Herschel} SPIRE 250/350/500 bands are caused by our implementation of the observational limits in these bands (see Sect.~\ref{Instruments.sec}), which heavily depends on the precise 2D distribution of the fluxes in the simulated images.

Overall, we conclude that the quality of the dust grid and the number of photons in our default setup are sufficient for our purposes.



\begin{figure*}
  \includegraphics[width=0.7\textwidth]{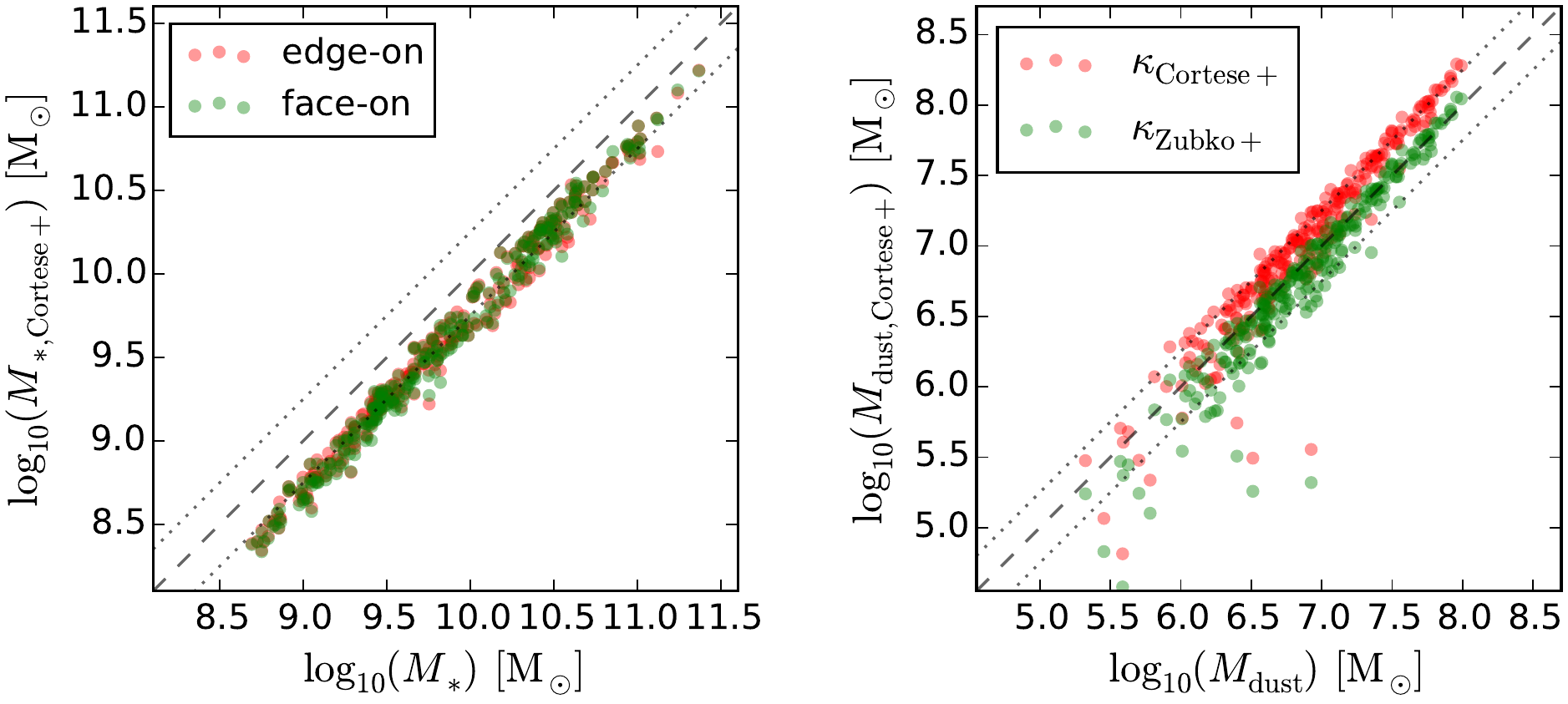}
  \caption{Comparison of stellar and dust mass derived from mock observations of the EAGLE galaxies in set \C\ (Table~\ref{GalaxySets.tab}) with the corresponding intrinsic properties. The EAGLE galaxies were post-processed using $f_\mathrm{dust}=0.3$ and $f_\mathrm{PDR}=0.1$. The dashed diagonal in each panel indicates the one-to-one relation; the dotted lines indicate $\pm$0.25 dex offsets. Left panel: stellar mass estimated through Eq.~(\ref{zibetti.eq}) following \citet{Cortese2012} and \citet{Zibetti2009} using edge-on (red) and face-on fluxes (green). Right panel: dust mass estimated through Eq.~(\ref{greybody.eq}) following \citet{Cortese2012} using $\beta=2$ and the two values of $\kappa_{350}$ defined in the text; galaxies with dust mass below $10^{4.75}\,\Msun$ are omitted.}
  \label{StellarAndDustMass.fig}
\end{figure*}


\section{Results and discussion}
\label{Results.sec}

In the following subsections we present results for EAGLE galaxies that were post-processed according to Eq.~(\ref{dustfromgas.eq}) with $f_\mathrm{dust}=0.3$ and $f_\mathrm{PDR}=0.1$. We will further justify these parameter values in Sect.~\ref{ParameterStudy.sec}.

\subsection{Intrinsic properties}
\label{IntrinsicProperties.sec}

Although our aim in this work is to evaluate mock observations of the EAGLE galaxies, it is instructive to briefly review some intrinsic properties, even if only to confirm that these values fall in the appropriate range. To this end, Figure~\ref{IntrinsicProperties.fig} shows selected intrinsic properties of the EAGLE galaxies analysed in this work, i.e.\ properties that can be calculated from the particles extracted from the snapshot without radiative transfer processing. Consistent with our selection criteria (Fig.~\ref{KBandHistogram.fig}), most high-mass and all early-type EAGLE galaxies are extracted from the Ref100 snapshot (red points). The remaining galaxies are extracted from the Recal25 snapshot (green points) or from the Ref25 snapshot (blue points) depending on the galaxy set under consideration (Table~\ref{GalaxySets.tab}).

The leftmost panel of Fig.~\ref{IntrinsicProperties.fig} plots specific star-formation rate (sSFR) versus stellar mass. \ch{As in Sect.~\ref{SelectingEAGLE.sec}, we can use the sSFR as a simple proxy for galaxy type, considering galaxies with a sSFR value below $10^{-11}~\mathrm{yr}^{-1}$ (indicated by the horizontal dashed line) to be early-type.} Comparing this diagram to, e.g., Fig.~8 of \citet{Kennicutt2012}, we conclude that both sSFR and stellar mass values are in the expected range, and we can clearly recognise a blue cloud of star-forming galaxies above the dashed line. The red sequence of quiescent galaxies below the dashed line is less prominent because our selection disfavours these galaxy types to reflect the HRS sample (see Sects.~\ref{HRS.sec} and \ref{SelectingEAGLE.sec}).

The middle panel of Fig.~\ref{IntrinsicProperties.fig} plots dust mass versus stellar mass. The dust mass is calculated by summing the result of Eq.~\ref{dustfromgas.eq} over all gas particles, using a dust-to-metal fraction $f_\mathrm{dust}=0.3$. Comparing this figure to, e.g., Fig.~16 of \citet{Bourne2012}, we conclude that these dust masses are within the expected range. The low dust masses for some of the high-stellar-mass (early-type) galaxies are also consistent with observations \citep{diSeregoAlighieri2013}.

The rightmost panel of the same figure plots the metallicity of the gas that contains dust versus stellar mass. The galaxies in our sample have a fairly high metallicity compared to observations \citep{Tremonti2004,Hughes2013,Zahid2014}. For example, the metallicities of the HRS galaxies shown in Fig.~4 of \citet{Hughes2013} do not exceed $\log_{10}(Z/\Zsun)=0.2$, assuming $12+\log_{10}(\mathrm{O}/\mathrm{H})_\odot=8.69$ \citep{Allende2001}. The high metallicities in our sample are, however, consistent with the mass-metallicity relation of the EAGLE galaxies reported in Fig.~13 of \citet{Schaye2015}. It is noted there that the Ref100 EAGLE simulation systematically over-predicts metallicity in the stellar mass range $M_*<10^{9.5}\,\Msun$. \ch{In addition to the uncertainties in both the normalization and the shape of the observed mass-metallicity relation, this discrepancy is most likely caused by the systematic uncertainties in the nucleosynthetic yields adopted in the EAGLE simulations.} It is also evident from the right panel of Fig.~\ref{IntrinsicProperties.fig} that galaxies in Recal25 tend to have lower metallicities than galaxies in Ref25, again consistent with the findings of \citet{Schaye2015}. \ch{Apparently, the stronger outflows in the Recal25 simulation reduce the metallicity of the ISM.} Because we use a constant dust-to-metal fraction, see Eq.~(\ref{dustfromgas.eq}), this leads to a slightly higher dust content for most Ref25 galaxies (middle panel of Fig.~\ref{IntrinsicProperties.fig}).


\begin{figure*}
  \includegraphics[width=\textwidth]{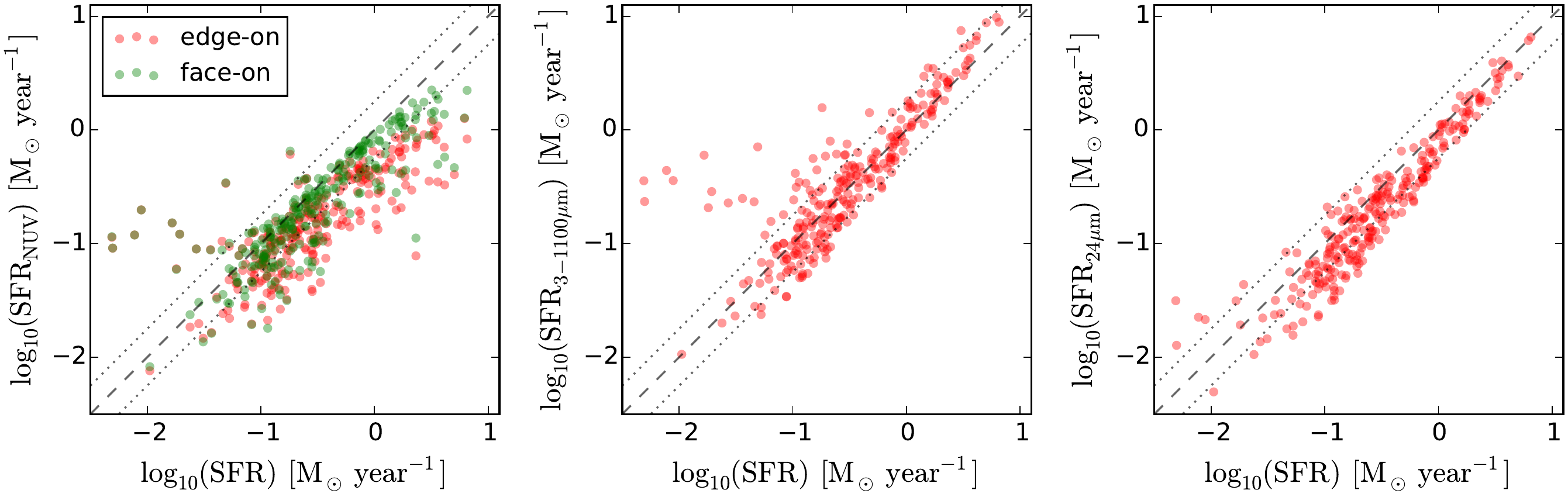}
  \caption{Comparison of three of the star-formation-rate (SFR) indicators summarised in Table~1 of \citet{Kennicutt2012}, calculated for mock observations of the EAGLE galaxies in set \C\ (Table~\ref{GalaxySets.tab}), to the intrinsic SFR provided in the public EAGLE database \citep{McAlpine2016}. The EAGLE galaxies were post-processed using $f_\mathrm{dust}=0.3$ and $f_\mathrm{PDR}=0.1$. The dashed diagonal in each panel indicates the one-to-one relation; the dotted lines indicate $\pm$0.25 dex offsets. Left panel: SFR based on GALEX NUV flux \citep{Hao2011,Murphy2011} using edge-on (red) and face-on fluxes (green). Middle panel: SFR based on integrated 3-1100~\micron\ flux \citep{Hao2011,Murphy2011}. Right panel: SFR based on \emph{Spitzer} MIPS 24~\micron\ flux \citep{Rieke2009}. In all panels, galaxies with intrinsic or inferred SFR below $10^{-2.5}\,\Msun\,\mathrm{year}^{-1}$ are omitted.}
  \label{StarFormationRate.fig}
\end{figure*}


\subsection{Inferred stellar and dust masses}
\label{StellarAndDustMass.sec}

We compare the stellar mass derived from mock observations of our EAGLE galaxies to the intrinsic stellar mass calculated by summing over all stellar particles. We mimic the procedure employed by \citet{Cortese2012}, determining the `mock' stellar mass $M_*$ from the $i$-band luminosity $L_i$ and the $g-i$ colour through
\begin{equation}
\log_{10} \frac{M_*}{\Msun} = \log_{10} \frac{L_i}{\mathrm{L}_{i,\odot}} + a + b \times (g-i),
\label{zibetti.eq}
\end{equation}
with coefficients $a=-0.963$ and $b=1.032$ taken from Table~B1 of \citet{Zibetti2009}. The result is shown in the left panel of Fig.~\ref{StellarAndDustMass.fig}. The mock observations of our EAGLE galaxies underestimate the stellar mass by about 0.25 dex, for both edge-on and face-on fluxes. The \citet{Zibetti2009} recipe assumes the \citet{Chabrier2003} initial mass function (IMF), as do the EAGLE simulations \citep[see Sect.~4.3 of][]{Schaye2015} and the SED templates we assign to stellar particles \citep[see our Sect.~\ref{SEDs.sec} and Sect.~2.3 of][]{Bruzual2003}, so there is no need to compensate for offsets between different IMFs in the model. However, the \citet{Zibetti2009} calibration of the stellar mass-to-light ratio relation was derived for resolved parts of galaxies. Several authors have proposed different values for the coefficients $a$ and $b$ \citep[e.g.,][]{Gallazzi2009,Taylor2010,Taylor2011,Baldry2012}, resulting in a systematic shift of up to 0.3 dex in the relation. In the following sections, we use the \citet{Zibetti2009} calibration because we will be confronting the mock observations with the results presented by \citet{Cortese2012}.

We now compare the dust mass derived from mock observations with the dust mass calculated by summing over all gas particles according to Eq.~(\ref{dustfromgas.eq}). Following \citet{Cortese2012} and many other authors, the flux $f_\nu$ emitted by a modified black body at the frequency $\nu$ can be written as
\begin{equation}
f_\nu = \frac{M_\mathrm{dust}}{d^2} \,\kappa_\nu\,B_\nu(T_\mathrm{dust})
\qquad \mathrm{with} \quad
\kappa_\nu=\kappa_{350}\left(\frac{\nu}{\nu_{350}}\right)^\beta,
\label{greybody.eq}
\end{equation}
where $M_\mathrm{dust}$ is the dust mass, $d$ is the distance, $B_\nu(T)$ is the Planck function, $T_\mathrm{dust}$ is the dust temperature, $\kappa_\nu$ is the dust mass absorption coefficient, assumed to depend on frequency through a power law with index $\beta$, and $\kappa_{350}$ is the dust mass absorption coefficient at a wavelength of $350~\micron$. \citet{Cortese2012} use the values $\beta=2$ and $\kappa_{350}=\kappa_{\mathrm{Cortese+}}=0.192~\mathrm{m}^2\,\mathrm{kg}^{-1}$. The \citet{Zubko2004} dust model used in this work (see Sect.~\ref{DustModel.sec}) has the same power-law index, $\beta=2$. However, the absorption coefficient $\kappa_{350}=\kappa_{\mathrm{Zubko+}}=0.330~\mathrm{m}^2\,\mathrm{kg}^{-1}$ differs substantially, causing a shift of 0.24 dex in the inferred dust mass. 

\citet{Cortese2012} use the three \emph{Herschel} SPIRE 250/350/500 fluxes to estimate the dust mass, employing a recipe presented in their Appendix B. The right panel of Fig.~\ref{StellarAndDustMass.fig} plots the dust mass estimates calculated from SKIRT fluxes for our EAGLE galaxies according to this recipe, using $\beta=2$ and the two values of $\kappa_{350}$ defined in the previous paragraph. When using $\kappa_{350}=\kappa_{\mathrm{Cortese+}}$, the \citet{Cortese2012} recipe overestimates the dust mass. With the $\kappa_{350}=\kappa_{\mathrm{Zubko+}}$ appropriate for our dust model, however, the estimates are fairly accurate, although there is significant scatter in the low mass range. Because we will be confronting our mock observations with the results presented by \citet{Cortese2012}, we use their dust mass recipe with $\kappa_{350}=\kappa_{\mathrm{Cortese+}}$ in the following sections.

\subsection{Inferred star-formation rates}
\label{SFRtracers.sec}

We compare in Fig.~\ref{StarFormationRate.fig} three of the star-formation-rate (SFR) indicators listed in Table~1 of \citet{Kennicutt2012}, calculated for mock observations of our EAGLE galaxies, to the intrinsic SFR provided in the public EAGLE database \citep{McAlpine2016}. The leftmost panel of Fig.~\ref{StarFormationRate.fig} shows the SFR based on the GALEX NUV flux \citep{Hao2011,Murphy2011} using edge-on (red) and face-on fluxes (green). At these short wavelengths, the edge-on fluxes suffer significantly more from dust extinction than the face-on fluxes, especially in more active galaxies, and thus yield a correspondingly lower SFR. However, even the indicator based on face-on fluxes slightly underestimates the SFR for most galaxies. For a small number of outliers, mostly in the lower SFR regime, the indicator substantially overestimates the SFR. These outliers are passive galaxies with a low dust content (the edge-on and face-on fluxes are equal so there is little extinction), where the NUV radiation emitted by the evolved star population is interpreted as a sign of star formation by the indicator. This so-called UV-upturn is also found in observations \citep[e.g.,][]{Brown1997,Brown2003}.

The middle panel of Fig.~\ref{StarFormationRate.fig} shows the SFR based on the integrated total infrared flux \citep[3-1100~\micron;][]{Hao2011,Murphy2011}. Because dust is mostly transparent to radiation at these wavelengths, the emission is isotropic and there is no need to compare edge-on and face-on fluxes. This indicator is fairly accurate, except for a number of outliers mostly in the lower SFR regime. In these cases, the emission from diffuse dust residing in the outskirts of those galaxies is interpreted by the indicator as a sign of star formation, while the dust is in fact being heated by an evolved star population. This phenomenon is also found in observations \citep{Bendo2015}.

The rightmost panel of Fig.~\ref{StarFormationRate.fig} shows the SFR based on the \emph{Spitzer} MIPS 24~\micron\ flux \citep{Rieke2009}. Except at the lowest SFRs, this indicator consistently underestimates the intrinsic SFR of our galaxies, which can be understood as follows. The EAGLE simulations do not model the cold ISM phase, and the adjustments made by our post-processing procedure have limitations as well. For example, our model contains isotropically emitting star-forming regions that may not represent the strong variations in the radiation field near star-forming regions sufficiently or accurately. As a consequence, at least some fraction of the diffuse dust in the simulated galaxies is heated insufficiently, resulting in a 24~\micron\ flux lower than observed. In addition, the flux in this wavelength range is very sensitive to the precise properties of the dust, and thus, the dust model used for the simulations \citep{Fanciullo2015}. For example, if the dust grain population contains a larger fraction of small grains, more grains will be stochastically heated to higher energy levels, shifting some of the dust emission to shorter wavelengths and into the 24~\micron\ band, while the total infrared flux remains unchanged.


\begin{figure*}
  \includegraphics[width=\textwidth]{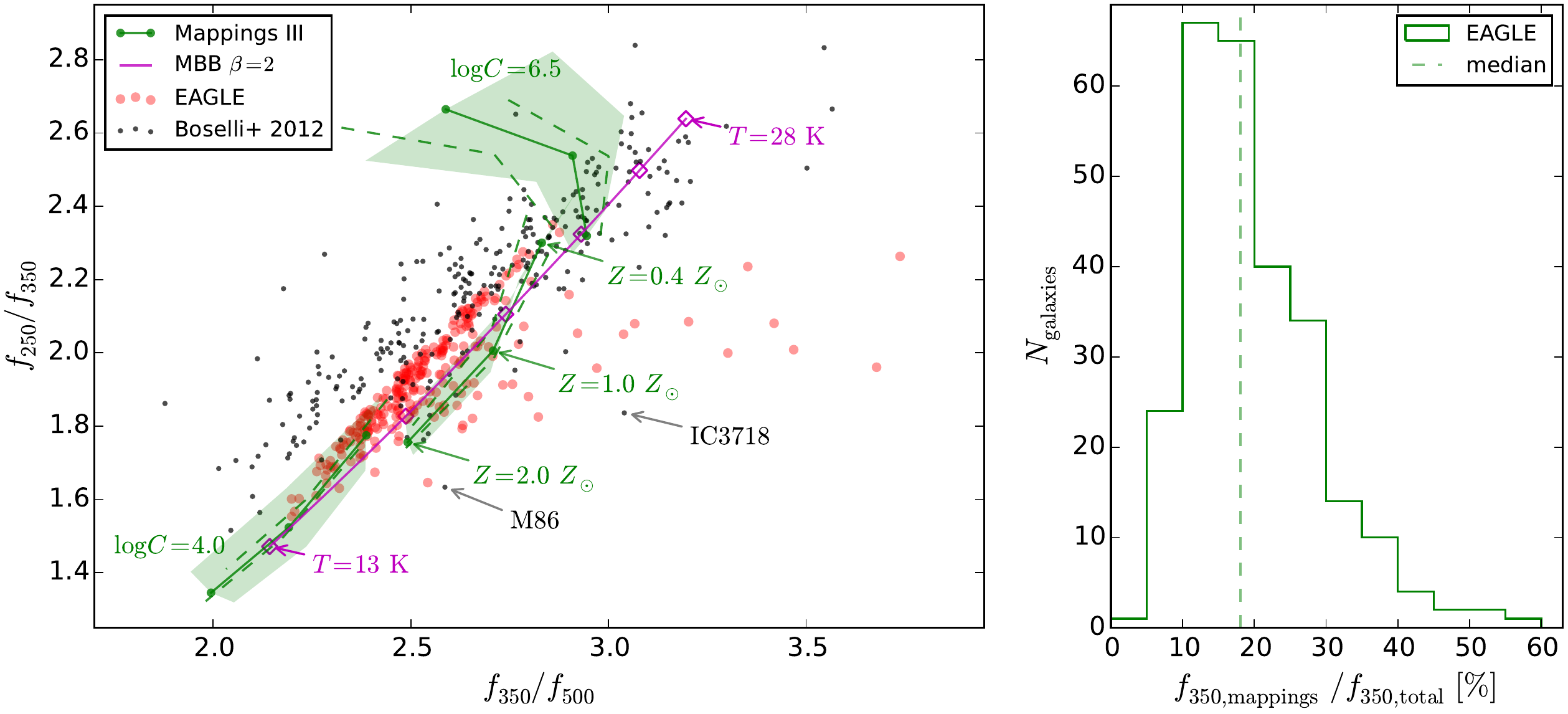} 
  \caption{\textsc{Left panel}: \emph{Herschel} SPIRE colour-colour relation $f_{250}/f_{350}$ versus $f_{350}/f_{500}$ for EAGLE galaxies (red points) from set \C\ (Table~\ref{GalaxySets.tab}) compared to HRS observations (black points) taken from \citet{Boselli2012} and \citet{Ciesla2012}. The EAGLE galaxies were post-processed using $f_\mathrm{dust}=0.3$ and $f_\mathrm{PDR}=0.1$. Galaxies for which one or more SPIRE fluxes are below the detection limit are omitted. The magenta curve traces a modified black body (MBB) with $\beta=2$ for temperatures ranging from 13~K to 28~K; the diamonds are spaced by 3~K. The green annotations indicate the flux ratios of the MAPPINGS III templates used to model star-forming regions for a range of input parameter values (see Table~\ref{Parameters.tab}). The solid lines show ratios for fixed $f_\mathrm{PDR}=0.1$ and for the extreme compactness values supported by the templates ($\log_{10} C=4.0,6.5$) plus an intermediate value ($\log_{10} C=5.25$). The dots on the solid lines show the ratios for varying metallicity ($Z=0.4,1,2 \times \Zsun$) with the given compactness. The dashed lines indicate the variation resulting from adjusting $f_\mathrm{PDR}$ to values of $0.05$ (higher temperature) or $0.15$ (lower temperature). The shaded areas indicate the variation resulting from adjusting the ambient pressure to the extreme values supported by the templates. \textsc{Right panel}: histogram of the MAPPINGS III contribution to the total flux at 350 \micron\ for the EAGLE galaxies shown in the left panel. The vertical dashed line indicates the median.}
  \label{Boselli.fig}
\end{figure*}

\begin{figure*}
  \includegraphics[width=\textwidth]{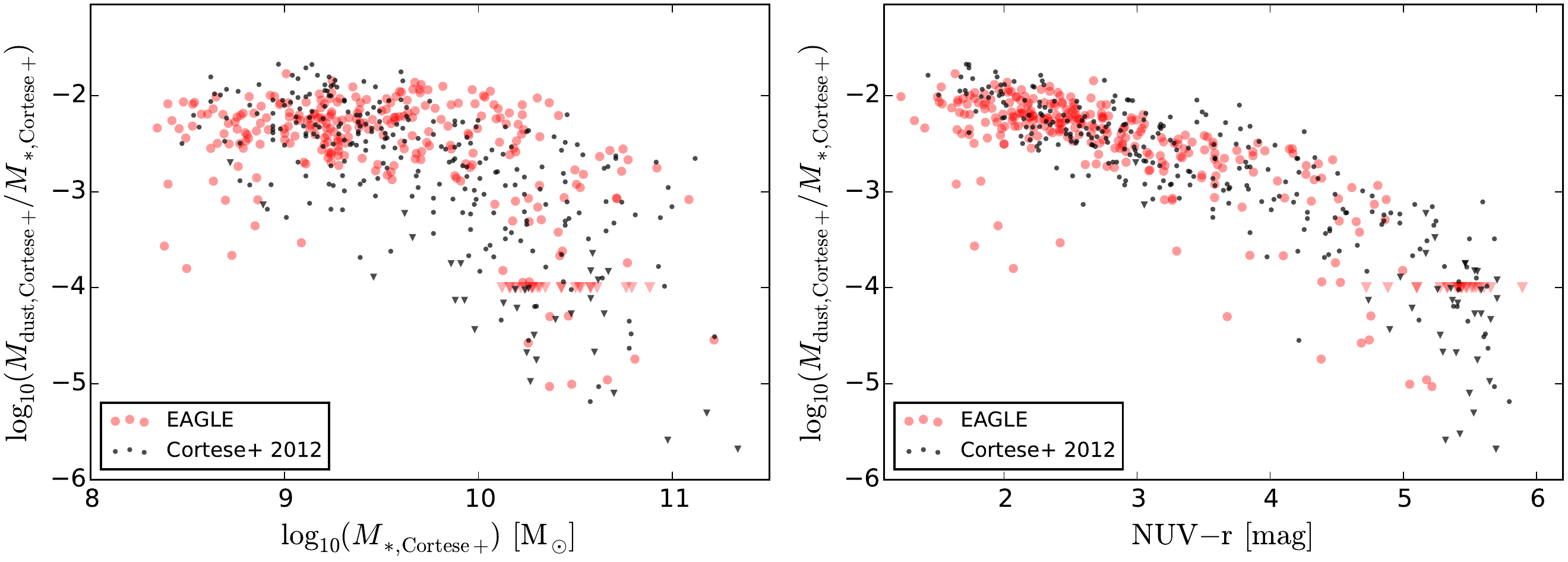}
  \caption{Dust scaling relations for EAGLE galaxies (red points) from set \C\ (Table~\ref{GalaxySets.tab}) compared to HRS observations (black points) taken from \citet{Cortese2012}. The EAGLE galaxies were post-processed using $f_\mathrm{dust}=0.3$ and $f_\mathrm{PDR}=0.1$, and all EAGLE data points are obtained through mock observations (see text). The left panel shows the dust-to-stellar mass ratio versus stellar mass. The right panel shows the dust-to-stellar mass ratio versus $\mathrm{NUV}-r$ colour. In both panels, galaxies for which one or more SPIRE fluxes are below the detection limit are plotted as filled triangles rather than circles. For the simulated galaxies, these non-detections are plotted at an arbitrary value of $M_\mathrm{dust}/M_*=10^{-4}$.}
  \label{Cortese.fig}
\end{figure*}

\begin{figure*}
  \includegraphics[width=\textwidth]{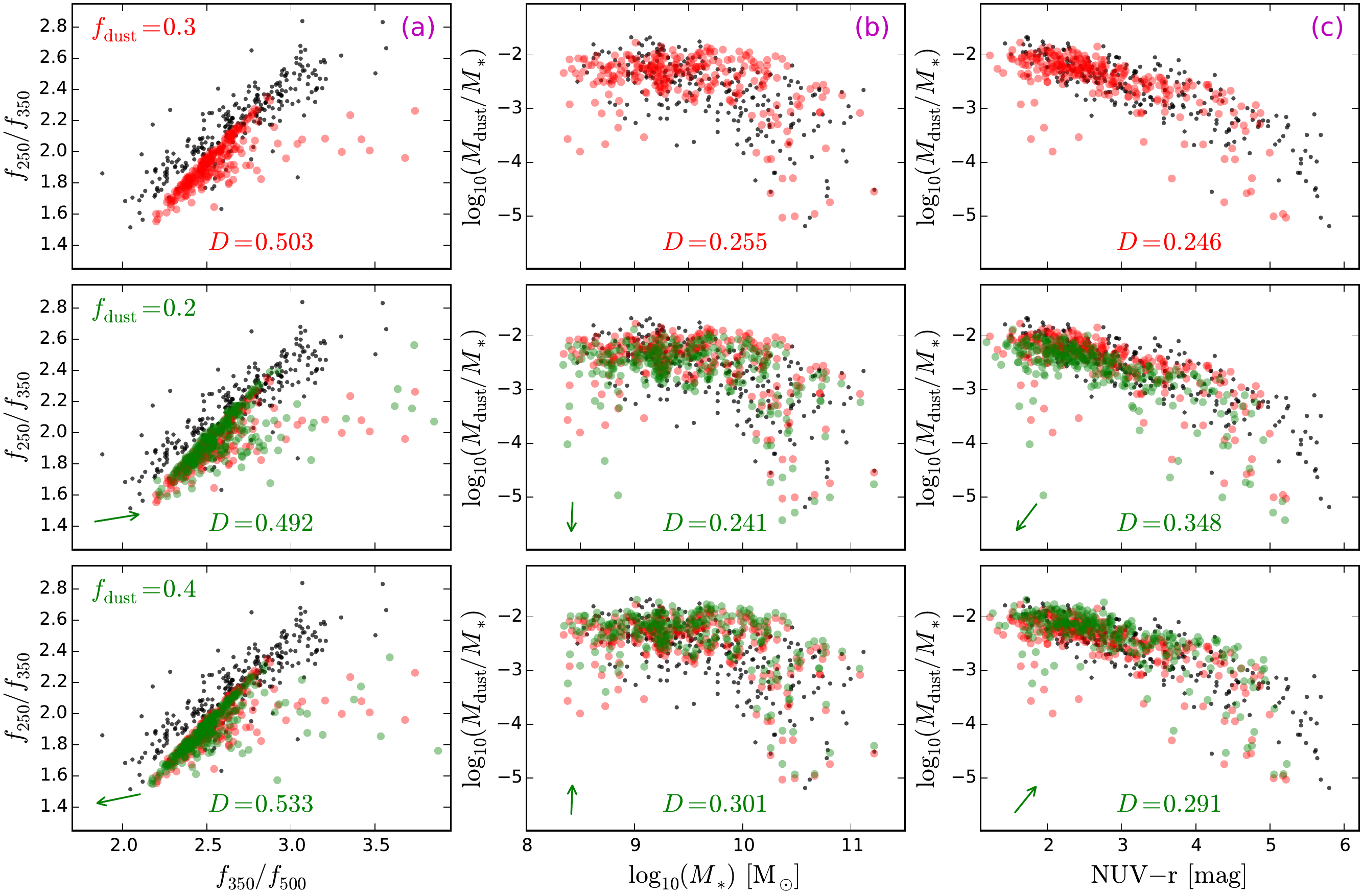}
  \caption{The scaling relations of Fig.~\ref{Boselli.fig}, shown in column (a), and Fig.~\ref{Cortese.fig}, shown in columns (b) and (c), for three values of the post-processing parameter $f_\mathrm{dust}$ (shown across the rows) with a fixed $f_\mathrm{PDR}=0.1$. The mock observations (red and green points) are overplotted on the HRS observations (black points) as before. The top row corresponds to the parameter settings of Figs.~\ref{Boselli.fig} and \ref{Cortese.fig}, i.e.\ $f_\mathrm{dust}=0.3$. These red points are repeated in the other rows for reference; the overplotted green points represent the mock observations with $f_\mathrm{dust}=0.2$ (middle row) and $f_\mathrm{dust}=0.4$ (bottom row). The green arrow in the lower left corner of these panels points in the direction of the average shift from red to green points (the length of the arrow is fixed). The $D$ value is a measure for the correspondence between the mock and HRS observations (smaller is better). Galaxies for which one or more SPIRE fluxes are below the detection limit are omitted from this figure and from the calculation of $D$.}
  \label{ParameterStudy_fdust.fig}
\end{figure*}

\begin{figure*}
  \includegraphics[width=\textwidth]{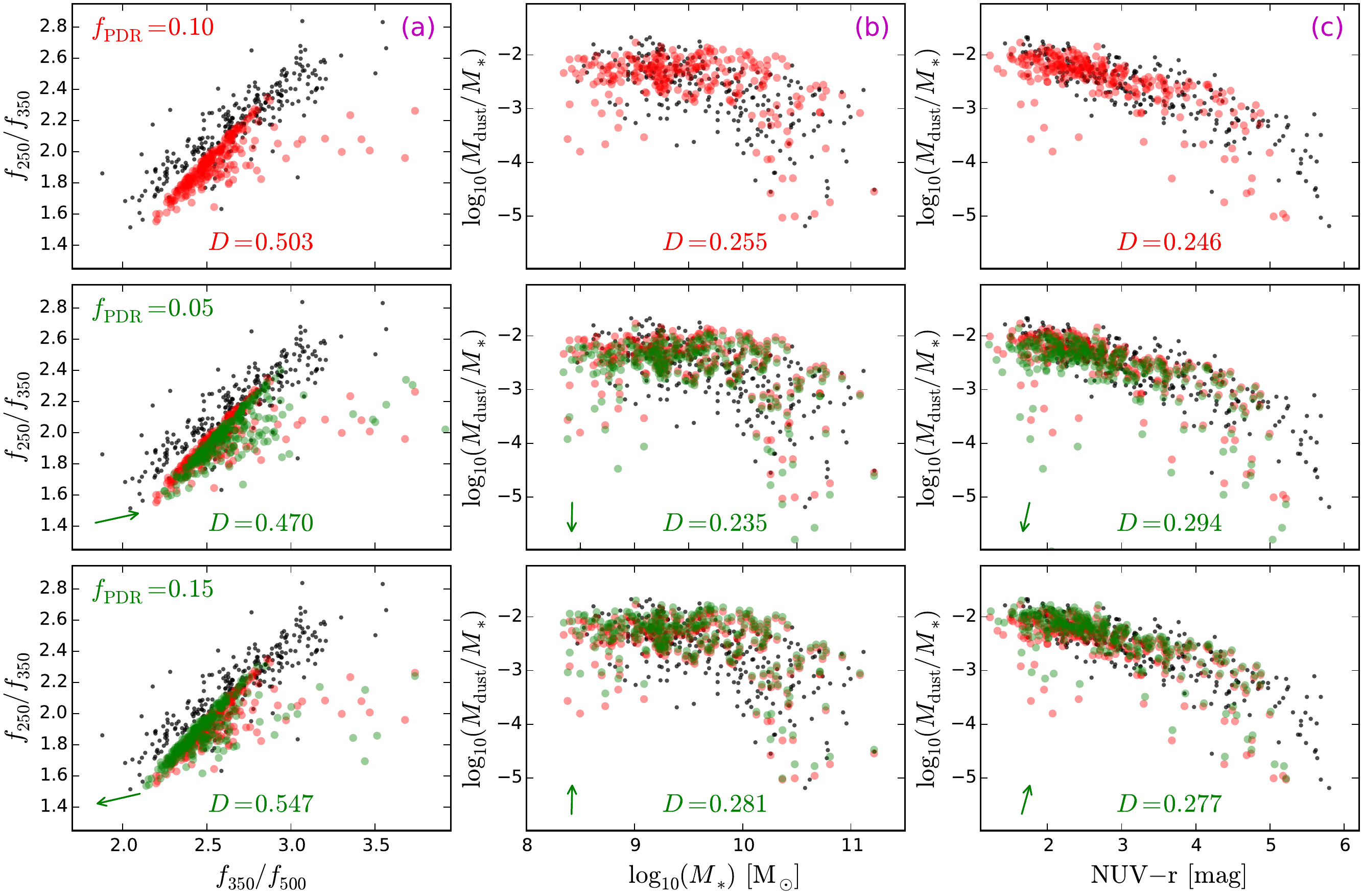}
  \caption{As Fig.~\ref{ParameterStudy_fdust.fig}, but now varying the post-processing parameter $f_\mathrm{PDR}$ with a fixed $f_\mathrm{dust}=0.3$. The top row again corresponds to the parameter settings of Figs.~\ref{Boselli.fig} and \ref{Cortese.fig}, i.e.\ $f_\mathrm{PDR}=0.1$. The overplotted green points in the other rows represent the mock observations with $f_\mathrm{PDR}=0.05$ (middle row) and $f_\mathrm{PDR}=0.15$ (bottom row).}
  \label{ParameterStudy_fpdr.fig}
\end{figure*}


\subsection{Dust scaling relations}
\label{DustScaling.sec}

In Fig.~\ref{Boselli.fig}, left panel, we compare mock observations of our EAGLE galaxies (red points) to HRS observations taken from \citet{Boselli2012} and \citet{Ciesla2012} (black points) for a submm colour-colour relation involving the SPIRE fluxes $f_{250}$, $f_{350}$, and $f_{500}$. These fluxes characterise the downwards slope of the dust continuum emission, and thus are sensitive to the cold dust contents. Smaller flux ratios $f_{250}/f_{350}$ and $f_{350}/f_{500}$ indicate a flatter slope of the dust emission curve and thus a larger contribution from colder dust. This is illustrated in the figure by the magenta curve, which traces the flux ratio relation for the emission of a modified black body (MBB, see Eq.~\ref{greybody.eq}) with $\beta=2$ (the value assumed by the dust model used in this work) for temperatures ranging from 13~K to 28~K. Data points away from this curve indicate contributions from dust at various temperatures, resulting in a superposition of MBB curves and thus a broader dust continuum emission peak.

The EAGLE and HRS scaling relations show similar slopes. However, the EAGLE flux ratios are generally smaller than the corresponding HRS flux ratios, i.e.\ the simulated points are shifted to the area of the plot indicating lower dust temperatures. The EAGLE galaxies thus show a larger contribution from very cold dust ($T\lesssim$~18~K), confirming our finding in Sect.~\ref{SFRtracers.sec} that part of the simulated dust is insufficiently heated, plausibly because of limitations in the sub-grid models of the simulations. It appears that, even with the sub-grid techniques in our procedures, our model does not fully capture the clumpiness of the dust distribution in galaxies, so that an insufficient amount of dust is irradiated by the strong radiation fields present within and near star-forming regions.

The EAGLE outliers to the right of the MBB curve are caused by the simulated observational limitations built into our procedure. This can be understood by noting that the observational limitations are more severe for longer wavelengths, and furthermore depend on the absolute flux level and its spatial distribution across the sky, so that the effect on the observed flux ratios is strongly nonlinear. In fact, in a version of the plot (not shown) using convolved but unlimited submm fluxes, i.e.\ skipping the operations described in the last paragraph of Sect.~\ref{Instruments.sec}, the EAGLE outliers move to the left of the MBB curve. We may thus surmise that the outlying HRS data points to the right of the MBB curve are similarly caused by observational limitations. The two labeled outliers, IC3718 and M86, are close to the SPIRE detection limit \citep{Boselli2010, Gomez2010}, supporting this assumption.

To help elucidate the contribution of the star-forming regions to the dust emission from the EAGLE galaxies, the green annotations in the left panel of Fig.~\ref{Boselli.fig} indicate the submm flux ratios of the MAPPINGS III templates used to model these regions for a range of input parameter values (see Table~\ref{Parameters.tab}). The effective dust temperature of an isolated MAPPINGS III star-forming region is mostly determined by the specified metallicity, $Z$, and compactness, $C$. The other parameters, including the covering fraction, $f_\mathrm{PDR}$, and the ambient pressure, $P$, have a much smaller impact. The higher-than-observed EAGLE metallicities may thus contribute to the lower dust temperatures. At the same time, the temperature increases rapidly with the \HII\ region's compactness. In fact, the temperature range $T\gtrsim$~22~K is reached only for the highest $C$ values supported by the MAPPINGS III templates. The compactness in turn increases with the \HII\ region mass (see Eq.~\ref{compactness.eq}). We sample these masses for our star-forming regions from a power-law distribution in the range $M\in[10^{2.8},10^6]\,\Msun$ (see Eq.~\ref{mass-distrib.eq}), while \citet{Groves2008} quote a range of $M\in[10^{3.5},10^7]\,\Msun$, almost an order of magnitude higher. Our lower masses, and thus lower $C$ values, might contribute to the apparent lack of warmer dust in our simulated results. These effects are limited, however, because for most galaxies the contribution of the MAPPINGS III templates to the total flux in the submm range is less than 25 per cent, as shown in the right panel of Fig.~\ref{Boselli.fig}.

In Fig.~\ref{Cortese.fig} we compare the dust scaling relations for mock observations of our EAGLE galaxies (red points) to HRS observations taken from \citet{Cortese2012} (black points). In order to make the comparison as unbiased as possible, rather than employing intrinsic properties obtained by summing over the smoothed particles, all EAGLE data points in Fig.~\ref{Cortese.fig} are derived from mock observations based on the SEDs generated by \SKIRT. Specifically, the stellar masses and dust masses are obtained following the recipes employed by \citet{Cortese2012} and summarised in Sect.~\ref{StellarAndDustMass.sec}. Galaxies for which one or more SPIRE fluxes are below the detection limit are plotted as filled triangles rather than circles. For the simulated galaxies, these non-detections are plotted at a fixed value of $M_\mathrm{dust}/M_*=10^{-4}$.

The dust-to-stellar mass ratio versus stellar mass scaling relation (left panel of Fig.~\ref{Cortese.fig}) is reproduced well, including the turn-off above $M_*=10^{10}\,\Msun$, although there is a slight positive offset in the dust-to-stellar mass ratio. The relation of dust-to-stellar mass ratio versus $\mathrm{NUV}-r$ colour (right panel of Fig.~\ref{Cortese.fig}) is also reproduced well, including the turn-off. As noted by \citet{Cortese2012}, the dust-to-stellar mass ratio anti-correlates strongly with stellar mass and with $\mathrm{NUV}-r$ colour, a proxy for sSFR \citep[e.g.,][]{Schiminovich2007}. The relations are remarkably similar to the scaling relation involving the \HI-to-stellar mass ratio \citep{Catinella2010, Cortese2011, Fabello2011}, suggesting that the dust and atomic hydrogen content of galaxies might be directly linked.

These results show that our EAGLE galaxies can indeed reproduce observed scaling relations based on FIR and submm fluxes. In the next section, we will present a quantitative measure for the correspondence between the simulations and data, and we will investigate the effect of changing the parameter values in our post-processing procedure.


\begin{figure*}
  \includegraphics[width=\textwidth]{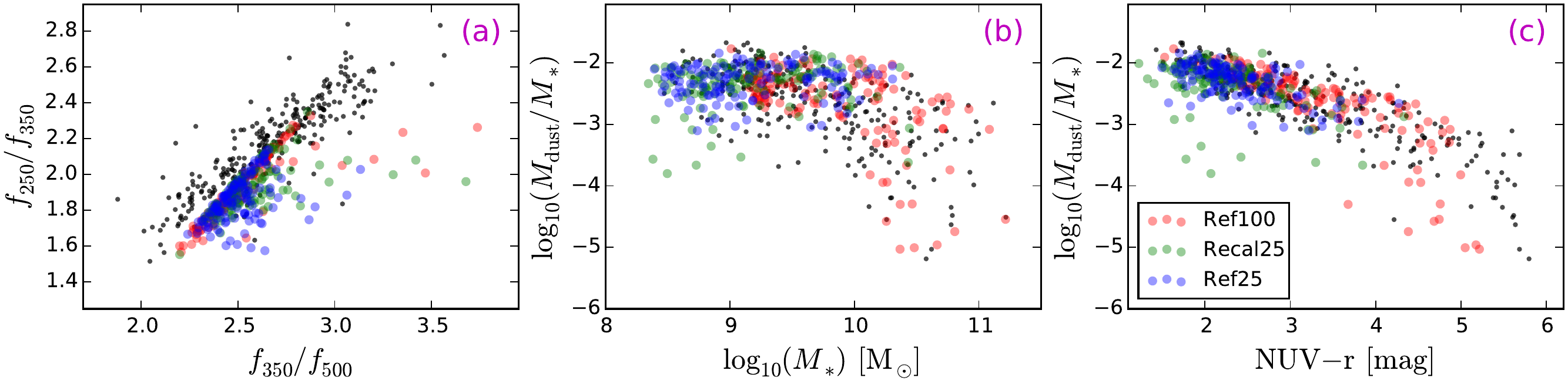}
  \caption{The scaling relations of Fig.~\ref{Boselli.fig}, shown in panel (a), and Fig.~\ref{Cortese.fig}, shown in panels (b) and (c), for all galaxies in sets \C\ and \F\ (see Table~\ref{GalaxySets.tab}). The points for our galaxies from snapshot Ref25 (blue) are plotted on top of those from Ref100 (red) and Recal25 (green) and the HRS observations (black). As in Figs.~\ref{Boselli.fig} and \ref{Cortese.fig}, the EAGLE galaxies were post-processed using $f_\mathrm{dust}=0.3$ and $f_\mathrm{PDR}=0.1$. Galaxies for which one or more SPIRE fluxes are below the detection limit are omitted from this plot.}
  \label{RefRecal.fig}
\end{figure*}

\begin{figure*}
  \includegraphics[width=\textwidth]{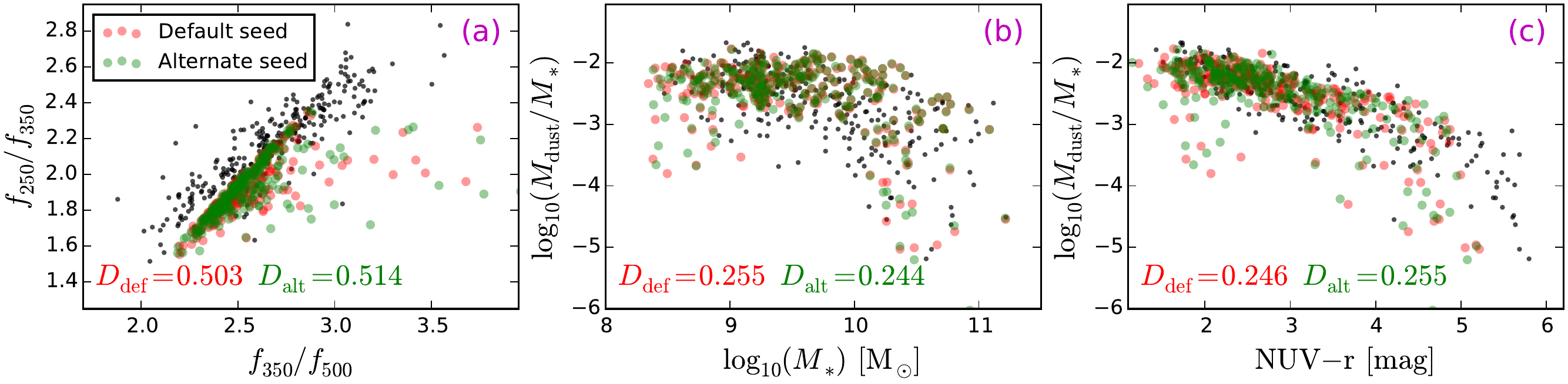}
  \caption{The scaling relations of Fig.~\ref{Boselli.fig}, shown in panel (a), and Fig.~\ref{Cortese.fig}, shown in panels (b) and (c), for the galaxies in set \C\ using $f_\mathrm{dust}=0.3$ and $f_\mathrm{PDR}=0.1$. The star-forming regions are re-sampled with the default (red) and an alternate (green) pseudo-random sequence to evaluate the effect of these random variations on the SKIRT input model. The $D$ value is a measure for the correspondence between the mock and HRS observations (smaller is better). Galaxies for which one or more SPIRE fluxes are below the detection limit are omitted from this plot.}
  \label{SamplingHII.fig}
\end{figure*}


\subsection{Parameter study}
\label{ParameterStudy.sec}

To quantify the correspondence between two sets of data points, such as the mock observations and the HRS data for one of the scaling relations shown in Figs.~\ref{Boselli.fig} and \ref{Cortese.fig}, we use a generalisation of the well-known Kolmogorov-Smirnov test \citep[K-S test,][]{Kolmogorov1933,Smirnov1948} to two-dimensional distributions. This generalisation is due to \citet{Peacock1983} and \citet{Fasano1987}, and its detailed implementation is described by \citet{Press2002}. The 2D K-S test computes a metric $D$ which can be interpreted as a measure of the `distance' between two sets of two-dimensional data points, with smaller $D$ values indicating better correspondence.

Figures~\ref{ParameterStudy_fdust.fig} and \ref{ParameterStudy_fpdr.fig} illustrate the effect of varying one of the post-processing parameters (respectively $f_\mathrm{dust}$ and $f_\mathrm{PDR}$) on the scaling relations of Figs.~\ref{Boselli.fig} and \ref{Cortese.fig}, i.e. relations between submm flux colours in column (a), dust-to-stellar mass ratio versus stellar mass in column (b), and dust-to-stellar mass ratio versus $\mathrm{NUV}-r$ colour in column (c). Each panel in these figures shows the K-S test $D$ value quantifying the discrepancy between the mock observations plotted in that panel and the underlying HRS data points. Because each column shows a particular scaling relation, $D$ values are comparable across the rows within each column, but not between columns.

Figure~\ref{ParameterStudy_fdust.fig} illustrates the effect of varying the dust-to-metal fraction $f_\mathrm{dust}$. According to Eq.~(\ref{dustfromgas.eq}), the diffuse dust mass in a simulated galaxy scales linearly with $f_\mathrm{dust}$ (the star-forming regions model a separate body of dust that is independent of $f_\mathrm{dust}$, see Sect.~\ref{SEDs.sec}). In columns (b) and (c) we indeed see that, with increasing dust-to-metal fraction, the `observed' dust-to-stellar mass ratio increases, and, for many galaxies, the $\mathrm{NUV}-r$ colour shifts red-ward because of the extinction caused by the extra dust. Also, as expected, there is very little effect on the `observed' stellar mass. The effect on the submm colours shown in column (a) is not pronounced, although there seems to be a slight shift towards flatter slopes, i.e.\ colder dust temperatures, with increasing $f_\mathrm{dust}$. This is because the larger body of diffuse dust is heated by the same stellar radiation to a lower average temperature.

Figure~\ref{ParameterStudy_fpdr.fig} illustrates the effect of varying the covering fraction $f_\mathrm{PDR}$ of the PDRs modelled in our simulations by the MAPPINGS III templates (see Sect.~\ref{SEDs.sec}). In column (a) we see a shift towards flatter slopes, i.e.\ colder dust temperatures, with increasing $f_\mathrm{PDR}$, caused by the more dispersed obscuration of the star-forming cores by the dust in the PDRs. In columns (b) and (c) we see an effect on the `observed' dust mass similar to the effect of varying the dust-to-metal fraction (Figure~\ref{ParameterStudy_fdust.fig}). This effect is caused by the additional dust emission modelled by the MAPPINGS III SEDs for increasing covering fractions. For example, it is clear from the top panel of Fig.~6 in \citet{Groves2008} that the continuum dust emission increases in addition to shifting to longer wavelengths. The apparent dust mass derived from a modified black body fitted to a MAPPINGS III SED increases by more than 80 per cent when $f_\mathrm{PDR}$ changes from $0.05$ to $0.1$, and by about another 50 per cent when $f_\mathrm{PDR}$ changes from $0.1$ to $0.15$, for fixed values of the other parameters (the precise percentages depend on the metallicity and other properties of the PDR). Varying the PDR covering fraction has only a small effect on the $\mathrm{NUV}-r$ colour because the dust mass is added in compact regions and does not contribute much to the overall extinction. In fact, in our simulations, the dust in the PDR regions does not contribute at all to the extinction of radiation originating outside of the PDR region, because each region is handled individually using a MAPPINGS III model.

While adjusting $f_\mathrm{dust}$ or $f_\mathrm{PDR}$ produces clear overall trends in these scaling relations, the effect varies substantially between individual galaxies, as is especially noticeable for some of the outliers. These differences are caused by varying degrees of extinction by diffuse dust and different levels of contribution from star-forming regions. Also, our emulation of the observational limitations, including instrument resolution and sensitivity, cause a nonlinear response of the `measured' flux to the actual submm emission of an EAGLE galaxy. This is especially true for galaxies with lower dust masses and thus less submm emission.

We used the $D$ value of the \citet{Cortese2012} scaling relation in column (c) of Figs.~\ref{ParameterStudy_fdust.fig} and \ref{ParameterStudy_fpdr.fig} to determine our standard set of post-processing parameter values, even though the $D$ value for the scaling relations in the other columns can be slightly more optimal for somewhat smaller amounts of dust (i.e. lower values of $f_\mathrm{dust}$ or $f_\mathrm{PDR}$). This leads to adopted values of $f_\mathrm{dust}=0.3$, fully compatible with the observed range of dust-to-metal fractions from $0.2$ to $0.4$ \citep{Dwek1998, Brinchmann2013, Zafar2013}, and $f_\mathrm{PDR}=0.1$, which is lower than the fiducial value of $0.2$ used by \citet{Jonsson2010}. It is worth noting once more in this context that the dust masses derived from our mock observations scale inversely with the assumed value of $\kappa_{350}$ (see Sect.~\ref{StellarAndDustMass.sec}). Using $\kappa_{\mathrm{Zubko+}}$ instead of $\kappa_{\mathrm{Cortese+}}$ would cause a 0.24 dex downward shift on the vertical axis in columns (b) and (c) of Figs.~\ref{ParameterStudy_fdust.fig} and \ref{ParameterStudy_fpdr.fig}. Adjusting our post-processing parameters to compensate for this shift would evidently affect the absolute dust masses assigned to our galaxies, without, however, changing the actual extinction levels.

We now briefly review the mock observations shown in Fig.~\ref{RefRecal.fig} for the galaxies extracted from the Ref25 snapshot (part of set \F, see Table~\ref{eaglesims.tab}). Based on the rightmost panel of Fig.~\ref{IntrinsicProperties.fig}, we already noted in Sect.~\ref{IntrinsicProperties.sec} that the Ref25 galaxies on average have a higher metallicity than the Recal25 galaxies, resulting in a larger dust mass. This effect is evident in columns (b) and (c) of Fig.~\ref{RefRecal.fig}, where the Ref25 galaxies (blue points) are positioned slightly higher, on average, compared to the Recal25 galaxies (green points). Overall the convergence between the lower-resolution Ref100 snapshot and the higher-resolution Ref/Recal25 snapshots is very good.

For the analysis presented so far, the star-forming region re-sampling procedure described in Sect.~\ref{Resampling.sec} was performed only once for each galaxy. In other words, for a given galaxy, exactly the same particle input sets were presented to SKIRT for all radiative transfer simulations of that galaxy. This approach has enabled us to focus on the effects of varying the values of $f_\mathrm{dust}$ and $f_\mathrm{PDR}$ in the radiative transfer model without interference from random changes in the input model. To verify that no biases were introduced by our specific instantiation of the star-forming regions, we reran the re-sampling procedure with a different pseudo-random sequence (i.e.\ a different seed). Fig.~\ref{SamplingHII.fig} shows the calculated scaling relations, using our standard values of $f_\mathrm{dust}=0.3$ and $f_\mathrm{PDR}=0.1$, for the default and alternate input models. While individual galaxies can differ substantially, especially the outliers, the overall results and conclusions remain unchanged.


\section{Conclusions}
\label{Conclusions.sec}

We calculated mock observations in the wavelength range from UV to submm for simulated galaxies extracted from the EAGLE suite of cosmological simulations using the radiative transfer code SKIRT. To help overcome some of the resolution limitations of the simulations, we employed sub-grid models for the star-forming regions and for the diffuse dust distribution. We also took special care to mimic the effects of instrumental properties and observational limitations when calculating band-integrated fluxes, which are important especially in submm bands.

To validate our method, and at the same time confront the properties of the simulated galaxies with observed galaxies, we selected a set of present-day EAGLE galaxies that matches the $K$-band luminosity distribution and overall morphological type classification (using the sSFR as a proxy) of the galaxies in the \emph{Herschel} Reference Survey (HRS), a volume-limited sample of about 300 normal galaxies in the Local Universe. We evaluated some intrinsic properties of our selected galaxies (Fig.~\ref{IntrinsicProperties.fig}), calculated by summing over the particles, and confirmed that the stellar masses, star-formation rates and dust masses fall in the expected range, while the average gas metallicities are above the metallicities observed in comparable galaxies.

We evaluated some relevant tracers by comparing the values derived from our mock observations to the corresponding intrinsic values (Figs.~\ref{StellarAndDustMass.fig} and \ref{StarFormationRate.fig}). We found that the \citet{Zibetti2009} calibration of stellar mass versus $g$ and $i$ band luminosity used by \citet{Cortese2012} underestimates stellar mass by about 0.25 dex, in line with the systematic uncertainty on the stellar mass-to-light ratio relation reported by other authors. Furthermore, the \citet{Cortese2012} recipe for deriving dust mass from the three SPIRE fluxes produces an offset that seems to be mostly caused by differences in the assumed dust absorption coefficient at 350~\micron. The star-formation indicators based on respectively NUV, 24~\micron, and integrated infrared fluxes perform fairly well, although the 24~\micron\  estimates are consistently low, most likely because some of the dust in our model is too cold.

We then studied dust scaling relations, including the $f_{250}/f_{350}$ versus $f_{350}/f_{500}$ submm colour-colour relation (Fig.~\ref{Boselli.fig}) and the dust-to-stellar mass ratio versus stellar mass and versus $\mathrm{NUV}-r$ colour relations (Fig.~\ref{Cortese.fig}), comparing the properties derived from mock observations of the EAGLE galaxies with those observed for HRS galaxies. Using our `standard' set of post-processing parameter values, we found good correspondence between the EAGLE and HRS scaling relations, with one important caveat. The submm colours indicate that the average temperature of the dust in our EAGLE galaxy models is lower than observed. We concluded that, even with the sub-grid techniques in our procedures, our model does not fully capture the clumpiness of the dust distribution in galaxies, so that an insufficient amount of dust is irradiated by the strong radiation fields present within and near star-forming regions.

We investigated the effects of varying the assumed dust-to-metal fraction, $f_\mathrm{dust}$ (Fig.~\ref{ParameterStudy_fdust.fig}), and PDR covering fraction, $f_\mathrm{PDR}$ (Fig.~\ref{ParameterStudy_fpdr.fig}), in our post-processing procedure. The first parameter controls the diffuse dust, while the latter controls the dust near young stellar populations. We found that the effects on the scaling relations are consistent with expectations, although it is hard to determine unambiguously optimal values for both parameters because of the partial degeneracy of the effects. We settled on $f_\mathrm{dust}=0.3$ and $f_\mathrm{PDR}=0.1$ as our standard values, noting that these values also depend on the properties of the dust in our model. These parameter values are compatible with observed values of $f_\mathrm{dust}$ \citep{Dwek1998, Brinchmann2013, Zafar2013} and values of $f_\mathrm{PDR}$ suggested by other authors \citep{Jonsson2010}.
 
In conclusion, our analysis has shown that, in spite of some limitations, the EAGLE simulations can reproduce infrared and submm observations through a physically motivated post-processing procedure. Furthermore, we have used the mock submm observations to constrain the important dust-related parameters in our method, leading to a consistent calculation of dust attenuation in the UV and optical domain by \citet{Trayford2016}. While we studied present-day galaxies in this work, our post-processing method is equally applicable to galaxies at higher redshifts, and could be readily adapted to other hydrodynamical simulations.

The method presented here opens many opportunities for future work. We plan to add infrared and submm fluxes for a large subset of the EAGLE galaxies at various redshifts to the public EAGLE database \citep{McAlpine2016} as a service to other researchers. We could study the morphology and structure of mock EAGLE galaxy images in various wavelength bands from UV to submm, comparing those properties with resolved observations in the same bands. It would also be instructive to post-process some of the more resolved galaxies in the zoom-in simulations \citep{Sawala2016, Oppenheimer2016} based on the EAGLE code.

In a more distant future, the astrophysical community will undoubtedly develop more advanced simulation techniques. Future simulations of galaxy evolution and assembly may use sub-grid recipes on a smaller scale to model a cold phase in the ISM, which will allow more accurate modeling of the clumpy dust structure during post-processing. Radiative transfer codes may include sub-grid models of star-forming regions that are connected to the overall RT model in a self-consistent manner, as opposed to employing `disconnected' SED templates. We hope that our current work will help inform such future efforts.

\section*{Acknowledgements}

The simulation data used in this work is available through collaboration with the authors. This work fits in the CHARM framework (Contemporary physical challenges in Heliospheric and AstRophysical Models), a phase VII Interuniversity Attraction Pole (IAP) programme organised by BELSPO, the BELgian federal Science Policy Office. This work used the DiRAC Data Centric system at Durham University, operated by the Institute for Computational Cosmology on behalf of the STFC DiRAC HPC Facility (www.dirac.ac.uk). This equipment was funded by BIS National E-infrastructure capital grant ST/K00042X/1, STFC capital grants ST/H008519/1 and ST/K00087X/1, STFC DiRAC Operations grant ST/K003267/1 and Durham University. DiRAC is part of the National E-Infrastructure. This research was supported in part by the European Research Council under the European Union's Seventh Framework Programme (FP7/2007-2013) / ERC Grant agreement 278594-GasAroundGalaxies.


\bibliographystyle{mnras}
\bibliography{eagledust}

\begin{thebibliography}{}
\makeatletter
\relax
\def\mn@urlcharsother{\let\do\@makeother \do\$\do\&\do\#\do\^\do\_\do\%\do\~}
\def\mn@doi{\begingroup\mn@urlcharsother \@ifnextchar [ {\mn@doi@}
  {\mn@doi@[]}}
\def\mn@doi@[#1]#2{\def\@tempa{#1}\ifx\@tempa\@empty \href
  {http://dx.doi.org/#2} {doi:#2}\else \href {http://dx.doi.org/#2} {#1}\fi
  \endgroup}
\def\mn@eprint#1#2{\mn@eprint@#1:#2::\@nil}
\def\mn@eprint@arXiv#1{\href {http://arxiv.org/abs/#1} {{\tt arXiv:#1}}}
\def\mn@eprint@dblp#1{\href {http://dblp.uni-trier.de/rec/bibtex/#1.xml}
  {dblp:#1}}
\def\mn@eprint@#1:#2:#3:#4\@nil{\def\@tempa {#1}\def\@tempb {#2}\def\@tempc
  {#3}\ifx \@tempc \@empty \let \@tempc \@tempb \let \@tempb \@tempa \fi \ifx
  \@tempb \@empty \def\@tempb {arXiv}\fi \@ifundefined
  {mn@eprint@\@tempb}{\@tempb:\@tempc}{\expandafter \expandafter \csname
  mn@eprint@\@tempb\endcsname \expandafter{\@tempc}}}

\bibitem[\protect\citeauthoryear{{Allende Prieto}, {Lambert}  \&
  {Asplund}}{{Allende Prieto} et~al.}{2001}]{Allende2001}
{Allende Prieto} C.,  {Lambert} D.~L.,   {Asplund} M.,  2001, \mn@doi [\apjl]
  {10.1086/322874}, \href {http://adsabs.harvard.edu/abs/2001ApJ...556L..63A}
  {556, L63}

\bibitem[\protect\citeauthoryear{{Altay} \& {Theuns}}{{Altay} \&
  {Theuns}}{2013}]{Altay2013}
{Altay} G.,  {Theuns} T.,  2013, \mn@doi [\mnras] {10.1093/mnras/stt1067},
  \href {http://adsabs.harvard.edu/abs/2013MNRAS.434..748A} {434, 748}

\bibitem[\protect\citeauthoryear{{Baes} \& {Camps}}{{Baes} \&
  {Camps}}{2015}]{Baes2015}
{Baes} M.,  {Camps} P.,  2015, \mn@doi [Astronomy and Computing]
  {10.1016/j.ascom.2015.05.006}, \href
  {http://adsabs.harvard.edu/abs/2015A%26C....12...33B} {12, 33}

\bibitem[\protect\citeauthoryear{{Baes} \& {Dejonghe}}{{Baes} \&
  {Dejonghe}}{2002}]{Baes2002}
{Baes} M.,  {Dejonghe} H.,  2002, \mn@doi [\mnras]
  {10.1046/j.1365-8711.2002.05641.x}, \href
  {http://adsabs.harvard.edu/abs/2002MNRAS.335..441B} {335, 441}

\bibitem[\protect\citeauthoryear{{Baes} et~al.,}{{Baes}
  et~al.}{2010}]{Baes2010}
{Baes} M.,  et~al., 2010, \mn@doi [\aap] {10.1051/0004-6361/201014644}, \href
  {http://adsabs.harvard.edu/abs/2010A%26A...518L..39B} {518, L39}

\bibitem[\protect\citeauthoryear{{Baes}, {Verstappen}, {De Looze}, {Fritz},
  {Saftly}, {Vidal P{\'e}rez}, {Stalevski}  \& {Valcke}}{{Baes}
  et~al.}{2011}]{Baes2011}
{Baes} M.,  {Verstappen} J.,  {De Looze} I.,  {Fritz} J.,  {Saftly} W.,  {Vidal
  P{\'e}rez} E.,  {Stalevski} M.,   {Valcke} S.,  2011, \mn@doi [\apjs]
  {10.1088/0067-0049/196/2/22}, \href
  {http://adsabs.harvard.edu/abs/2011ApJS..196...22B} {196, 22}

\bibitem[\protect\citeauthoryear{{Bah{\'e}} et~al.,}{{Bah{\'e}}
  et~al.}{2016}]{Bahe2016}
{Bah{\'e}} Y.~M.,  et~al., 2016, \mn@doi [\mnras] {10.1093/mnras/stv2674},
  \href {http://adsabs.harvard.edu/abs/2016MNRAS.456.1115B} {456, 1115}

\bibitem[\protect\citeauthoryear{{Baldry} et~al.,}{{Baldry}
  et~al.}{2012}]{Baldry2012}
{Baldry} I.~K.,  et~al., 2012, \mn@doi [\mnras]
  {10.1111/j.1365-2966.2012.20340.x}, \href
  {http://adsabs.harvard.edu/abs/2012MNRAS.421..621B} {421, 621}

\bibitem[\protect\citeauthoryear{{Bendo} et~al.,}{{Bendo}
  et~al.}{2015}]{Bendo2015}
{Bendo} G.~J.,  et~al., 2015, \mn@doi [\mnras] {10.1093/mnras/stu1841}, \href
  {http://adsabs.harvard.edu/abs/2015MNRAS.448..135B} {448, 135}

\bibitem[\protect\citeauthoryear{{Boselli} et~al.,}{{Boselli}
  et~al.}{2010}]{Boselli2010}
{Boselli} A.,  et~al., 2010, \mn@doi [\pasp] {10.1086/651535}, \href
  {http://adsabs.harvard.edu/abs/2010PASP..122..261B} {122, 261}

\bibitem[\protect\citeauthoryear{{Boselli} et~al.,}{{Boselli}
  et~al.}{2012}]{Boselli2012}
{Boselli} A.,  et~al., 2012, \mn@doi [\aap] {10.1051/0004-6361/201118602},
  \href {http://adsabs.harvard.edu/abs/2012A%26A...540A..54B} {540, A54}

\bibitem[\protect\citeauthoryear{{Bourne} et~al.,}{{Bourne}
  et~al.}{2012}]{Bourne2012}
{Bourne} N.,  et~al., 2012, \mn@doi [\mnras]
  {10.1111/j.1365-2966.2012.20528.x}, \href
  {http://adsabs.harvard.edu/abs/2012MNRAS.421.3027B} {421, 3027}

\bibitem[\protect\citeauthoryear{{Brinchmann}, {Charlot}, {Kauffmann},
  {Heckman}, {White}  \& {Tremonti}}{{Brinchmann}
  et~al.}{2013}]{Brinchmann2013}
{Brinchmann} J.,  {Charlot} S.,  {Kauffmann} G.,  {Heckman} T.,  {White}
  S.~D.~M.,   {Tremonti} C.,  2013, \mn@doi [\mnras] {10.1093/mnras/stt551},
  \href {http://adsabs.harvard.edu/abs/2013MNRAS.432.2112B} {432, 2112}

\bibitem[\protect\citeauthoryear{{Brown}, {Ferguson}, {Davidsen}  \&
  {Dorman}}{{Brown} et~al.}{1997}]{Brown1997}
{Brown} T.~M.,  {Ferguson} H.~C.,  {Davidsen} A.~F.,   {Dorman} B.,  1997,
  \apj, \href {http://adsabs.harvard.edu/abs/1997ApJ...482..685B} {482, 685}

\bibitem[\protect\citeauthoryear{{Brown}, {Ferguson}, {Smith}, {Bowers},
  {Kimble}, {Renzini}  \& {Rich}}{{Brown} et~al.}{2003}]{Brown2003}
{Brown} T.~M.,  {Ferguson} H.~C.,  {Smith} E.,  {Bowers} C.~W.,  {Kimble}
  R.~A.,  {Renzini} A.,   {Rich} R.~M.,  2003, \mn@doi [\apjl]
  {10.1086/374035}, \href {http://adsabs.harvard.edu/abs/2003ApJ...584L..69B}
  {584, L69}

\bibitem[\protect\citeauthoryear{{Bruzual} \& {Charlot}}{{Bruzual} \&
  {Charlot}}{2003}]{Bruzual2003}
{Bruzual} G.,  {Charlot} S.,  2003, \mn@doi [\mnras]
  {10.1046/j.1365-8711.2003.06897.x}, \href
  {http://adsabs.harvard.edu/abs/2003MNRAS.344.1000B} {344, 1000}

\bibitem[\protect\citeauthoryear{{Byun}, {Freeman}  \& {Kylafis}}{{Byun}
  et~al.}{1994}]{Byun1994}
{Byun} Y.~I.,  {Freeman} K.~C.,   {Kylafis} N.~D.,  1994, \mn@doi [\apj]
  {10.1086/174553}, \href {http://adsabs.harvard.edu/abs/1994ApJ...432..114B}
  {432, 114}

\bibitem[\protect\citeauthoryear{{Camps} \& {Baes}}{{Camps} \&
  {Baes}}{2015}]{Camps2015a}
{Camps} P.,  {Baes} M.,  2015, \mn@doi [Astronomy and Computing]
  {10.1016/j.ascom.2014.10.004}, \href
  {http://adsabs.harvard.edu/abs/2015A%26C.....9...20C} {9, 20}

\bibitem[\protect\citeauthoryear{{Camps}, {Baes}  \& {Saftly}}{{Camps}
  et~al.}{2013}]{Camps2013}
{Camps} P.,  {Baes} M.,   {Saftly} W.,  2013, \mn@doi [\aap]
  {10.1051/0004-6361/201322281}, \href
  {http://adsabs.harvard.edu/abs/2013A%26A...560A..35C} {560, A35}

\bibitem[\protect\citeauthoryear{{Camps} et~al.,}{{Camps}
  et~al.}{2015}]{Camps2015b}
{Camps} P.,  et~al., 2015, \mn@doi [\aap] {10.1051/0004-6361/201525998}, \href
  {http://adsabs.harvard.edu/abs/2015A%26A...580A..87C} {580, A87}

\bibitem[\protect\citeauthoryear{{Catinella} et~al.,}{{Catinella}
  et~al.}{2010}]{Catinella2010}
{Catinella} B.,  et~al., 2010, \mn@doi [\mnras]
  {10.1111/j.1365-2966.2009.16180.x}, \href
  {http://adsabs.harvard.edu/abs/2010MNRAS.403..683C} {403, 683}

\bibitem[\protect\citeauthoryear{{Chabrier}}{{Chabrier}}{2003}]{Chabrier2003}
{Chabrier} G.,  2003, \mn@doi [\apjl] {10.1086/374879}, \href
  {http://adsabs.harvard.edu/abs/2003ApJ...586L.133C} {586, L133}

\bibitem[\protect\citeauthoryear{{Ciesla} et~al.,}{{Ciesla}
  et~al.}{2012}]{Ciesla2012}
{Ciesla} L.,  et~al., 2012, \mn@doi [\aap] {10.1051/0004-6361/201219216}, \href
  {http://adsabs.harvard.edu/abs/2012A%26A...543A.161C} {543, A161}

\bibitem[\protect\citeauthoryear{{Cohen}, {Wheaton}  \& {Megeath}}{{Cohen}
  et~al.}{2003}]{Cohen2003}
{Cohen} M.,  {Wheaton} W.~A.,   {Megeath} S.~T.,  2003, \mn@doi [\aj]
  {10.1086/376474}, \href {http://adsabs.harvard.edu/abs/2003AJ....126.1090C}
  {126, 1090}

\bibitem[\protect\citeauthoryear{{Cortese}, {Catinella}, {Boissier}, {Boselli}
  \& {Heinis}}{{Cortese} et~al.}{2011}]{Cortese2011}
{Cortese} L.,  {Catinella} B.,  {Boissier} S.,  {Boselli} A.,   {Heinis} S.,
  2011, \mn@doi [\mnras] {10.1111/j.1365-2966.2011.18822.x}, \href
  {http://adsabs.harvard.edu/abs/2011MNRAS.415.1797C} {415, 1797}

\bibitem[\protect\citeauthoryear{{Cortese} et~al.,}{{Cortese}
  et~al.}{2012}]{Cortese2012}
{Cortese} L.,  et~al., 2012, \mn@doi [\aap] {10.1051/0004-6361/201118499},
  \href {http://adsabs.harvard.edu/abs/2012A%26A...540A..52C} {540, A52}

\bibitem[\protect\citeauthoryear{{Crain} et~al.,}{{Crain}
  et~al.}{2015}]{Crain2015}
{Crain} R.~A.,  et~al., 2015, \mn@doi [\mnras] {10.1093/mnras/stv725}, \href
  {http://adsabs.harvard.edu/abs/2015MNRAS.450.1937C} {450, 1937}

\bibitem[\protect\citeauthoryear{{Dalla Vecchia} \& {Schaye}}{{Dalla Vecchia}
  \& {Schaye}}{2012}]{DallaVecchia2012}
{Dalla Vecchia} C.,  {Schaye} J.,  2012, \mn@doi [\mnras]
  {10.1111/j.1365-2966.2012.21704.x}, \href
  {http://adsabs.harvard.edu/abs/2012MNRAS.426..140D} {426, 140}

\bibitem[\protect\citeauthoryear{{De Geyter}, {Baes}, {Camps}, {Fritz}, {De
  Looze}, {Hughes}, {Viaene}  \& {Gentile}}{{De Geyter}
  et~al.}{2014}]{DeGeyter2014}
{De Geyter} G.,  {Baes} M.,  {Camps} P.,  {Fritz} J.,  {De Looze} I.,  {Hughes}
  T.~M.,  {Viaene} S.,   {Gentile} G.,  2014, \mn@doi [\mnras]
  {10.1093/mnras/stu612}, \href
  {http://adsabs.harvard.edu/abs/2014MNRAS.441..869D} {441, 869}

\bibitem[\protect\citeauthoryear{{De Looze}, {Baes}, {Fritz}  \&
  {Verstappen}}{{De Looze} et~al.}{2012}]{DeLooze2012}
{De Looze} I.,  {Baes} M.,  {Fritz} J.,   {Verstappen} J.,  2012, \mn@doi
  [\mnras] {10.1111/j.1365-2966.2011.19759.x}, \href
  {http://adsabs.harvard.edu/abs/2012MNRAS.419..895D} {419, 895}

\bibitem[\protect\citeauthoryear{{Deschamps}, {Braun}, {Jorissen}, {Siess},
  {Baes}  \& {Camps}}{{Deschamps} et~al.}{2015}]{Deschamps2015}
{Deschamps} R.,  {Braun} K.,  {Jorissen} A.,  {Siess} L.,  {Baes} M.,   {Camps}
  P.,  2015, \mn@doi [\aap] {10.1051/0004-6361/201424772}, \href
  {http://adsabs.harvard.edu/abs/2015A%26A...577A..55D} {577, A55}

\bibitem[\protect\citeauthoryear{{Disney}, {Davies}  \& {Phillipps}}{{Disney}
  et~al.}{1989}]{Disney1989}
{Disney} M.,  {Davies} J.,   {Phillipps} S.,  1989, \mnras, \href
  {http://adsabs.harvard.edu/abs/1989MNRAS.239..939D} {239, 939}

\bibitem[\protect\citeauthoryear{{Doi} et~al.,}{{Doi} et~al.}{2010}]{Doi2010}
{Doi} M.,  et~al., 2010, \mn@doi [\aj] {10.1088/0004-6256/139/4/1628}, \href
  {http://adsabs.harvard.edu/abs/2010AJ....139.1628D} {139, 1628}

\bibitem[\protect\citeauthoryear{{Dolag}, {Borgani}, {Murante}  \&
  {Springel}}{{Dolag} et~al.}{2009}]{Dolag2009b}
{Dolag} K.,  {Borgani} S.,  {Murante} G.,   {Springel} V.,  2009, \mn@doi
  [\mnras] {10.1111/j.1365-2966.2009.15034.x}, \href
  {http://adsabs.harvard.edu/abs/2009MNRAS.399..497D} {399, 497}

\bibitem[\protect\citeauthoryear{{Dom{\'{\i}}nguez-Tenreiro}, {Obreja},
  {Granato}, {Schurer}, {Alpresa}, {Silva}, {Brook}  \&
  {Serna}}{{Dom{\'{\i}}nguez-Tenreiro} et~al.}{2014}]{Dominguez-Tenreiro2014}
{Dom{\'{\i}}nguez-Tenreiro} R.,  {Obreja} A.,  {Granato} G.~L.,  {Schurer} A.,
  {Alpresa} P.,  {Silva} L.,  {Brook} C.~B.,   {Serna} A.,  2014, \mn@doi
  [\mnras] {10.1093/mnras/stu240}, \href
  {http://adsabs.harvard.edu/abs/2014MNRAS.439.3868D} {439, 3868}

\bibitem[\protect\citeauthoryear{{Dopita} et~al.,}{{Dopita}
  et~al.}{2005}]{Dopita2005}
{Dopita} M.~A.,  et~al., 2005, \mn@doi [\apj] {10.1086/423948}, \href
  {http://adsabs.harvard.edu/abs/2005ApJ...619..755D} {619, 755}

\bibitem[\protect\citeauthoryear{{Draine} \& {Lee}}{{Draine} \&
  {Lee}}{1984}]{Draine1984}
{Draine} B.~T.,  {Lee} H.~M.,  1984, \mn@doi [\apj] {10.1086/162480}, \href
  {http://adsabs.harvard.edu/abs/1984ApJ...285...89D} {285, 89}

\bibitem[\protect\citeauthoryear{{Draine} \& {Li}}{{Draine} \&
  {Li}}{2001}]{Draine2001}
{Draine} B.~T.,  {Li} A.,  2001, \mn@doi [\apj] {10.1086/320227}, \href
  {http://adsabs.harvard.edu/abs/2001ApJ...551..807D} {551, 807}

\bibitem[\protect\citeauthoryear{{Dwek}}{{Dwek}}{1998}]{Dwek1998}
{Dwek} E.,  1998, \mn@doi [\apj] {10.1086/305829}, \href
  {http://adsabs.harvard.edu/abs/1998ApJ...501..643D} {501, 643}

\bibitem[\protect\citeauthoryear{{Fabello}, {Catinella}, {Giovanelli},
  {Kauffmann}, {Haynes}, {Heckman}  \& {Schiminovich}}{{Fabello}
  et~al.}{2011}]{Fabello2011}
{Fabello} S.,  {Catinella} B.,  {Giovanelli} R.,  {Kauffmann} G.,  {Haynes}
  M.~P.,  {Heckman} T.~M.,   {Schiminovich} D.,  2011, \mn@doi [\mnras]
  {10.1111/j.1365-2966.2010.17742.x}, \href
  {http://adsabs.harvard.edu/abs/2011MNRAS.411..993F} {411, 993}

\bibitem[\protect\citeauthoryear{{Fanciullo}, {Guillet}, {Aniano}, {Jones},
  {Ysard}, {Miville-Desch{\^e}nes}, {Boulanger}  \& {K{\"o}hler}}{{Fanciullo}
  et~al.}{2015}]{Fanciullo2015}
{Fanciullo} L.,  {Guillet} V.,  {Aniano} G.,  {Jones} A.~P.,  {Ysard} N.,
  {Miville-Desch{\^e}nes} M.-A.,  {Boulanger} F.,   {K{\"o}hler} M.,  2015,
  \mn@doi [\aap] {10.1051/0004-6361/201525677}, \href
  {http://adsabs.harvard.edu/abs/2015A%26A...580A.136F} {580, A136}

\bibitem[\protect\citeauthoryear{{Fasano} \& {Franceschini}}{{Fasano} \&
  {Franceschini}}{1987}]{Fasano1987}
{Fasano} G.,  {Franceschini} A.,  1987, \mn@doi [\mnras]
  {10.1093/mnras/225.1.155}, \href
  {http://adsabs.harvard.edu/abs/1987MNRAS.225..155F} {225, 155}

\bibitem[\protect\citeauthoryear{{Furlong} et~al.,}{{Furlong}
  et~al.}{2015}]{Furlong2015}
{Furlong} M.,  et~al., 2015, \mn@doi [\mnras] {10.1093/mnras/stv852}, \href
  {http://adsabs.harvard.edu/abs/2015MNRAS.450.4486F} {450, 4486}

\bibitem[\protect\citeauthoryear{{Gallazzi} \& {Bell}}{{Gallazzi} \&
  {Bell}}{2009}]{Gallazzi2009}
{Gallazzi} A.,  {Bell} E.~F.,  2009, \mn@doi [\apjs]
  {10.1088/0067-0049/185/2/253}, \href
  {http://adsabs.harvard.edu/abs/2009ApJS..185..253G} {185, 253}

\bibitem[\protect\citeauthoryear{{Gomez} et~al.,}{{Gomez}
  et~al.}{2010}]{Gomez2010}
{Gomez} H.~L.,  et~al., 2010, \mn@doi [\aap] {10.1051/0004-6361/201014530},
  \href {http://adsabs.harvard.edu/abs/2010A%26A...518L..45G} {518, L45}

\bibitem[\protect\citeauthoryear{{Griffin} et~al.,}{{Griffin}
  et~al.}{2010}]{Griffin2010}
{Griffin} M.~J.,  et~al., 2010, \mn@doi [\aap] {10.1051/0004-6361/201014519},
  \href {http://adsabs.harvard.edu/abs/2010A%26A...518L...3G} {518, L3}

\bibitem[\protect\citeauthoryear{{Groves}, {Dopita}, {Sutherland}, {Kewley},
  {Fischera}, {Leitherer}, {Brandl}  \& {van Breugel}}{{Groves}
  et~al.}{2008}]{Groves2008}
{Groves} B.,  {Dopita} M.~A.,  {Sutherland} R.~S.,  {Kewley} L.~J.,  {Fischera}
  J.,  {Leitherer} C.,  {Brandl} B.,   {van Breugel} W.,  2008, \mn@doi [\apjs]
  {10.1086/528711}, \href {http://adsabs.harvard.edu/abs/2008ApJS..176..438G}
  {176, 438}

\bibitem[\protect\citeauthoryear{{Guhathakurta} \& {Draine}}{{Guhathakurta} \&
  {Draine}}{1989}]{Guhathakurta1989}
{Guhathakurta} P.,  {Draine} B.~T.,  1989, \mn@doi [\apj] {10.1086/167899},
  \href {http://adsabs.harvard.edu/abs/1989ApJ...345..230G} {345, 230}

\bibitem[\protect\citeauthoryear{{Guidi}, {Scannapieco}  \& {Walcher}}{{Guidi}
  et~al.}{2015}]{Guidi2015}
{Guidi} G.,  {Scannapieco} C.,   {Walcher} C.~J.,  2015, \mn@doi [\mnras]
  {10.1093/mnras/stv2050}, \href
  {http://adsabs.harvard.edu/abs/2015MNRAS.454.2381G} {454, 2381}

\bibitem[\protect\citeauthoryear{{Haardt} \& {Madau}}{{Haardt} \&
  {Madau}}{2001}]{Haardt2001}
{Haardt} F.,  {Madau} P.,  2001, in {Neumann} D.~M.,  {Tran} J.~T.~V.,  eds,
  Clusters of Galaxies and the High Redshift Universe Observed in X-rays.
  (\mn@eprint {} {astro-ph/0106018})

\bibitem[\protect\citeauthoryear{{Hao}, {Kennicutt}, {Johnson}, {Calzetti},
  {Dale}  \& {Moustakas}}{{Hao} et~al.}{2011}]{Hao2011}
{Hao} C.-N.,  {Kennicutt} R.~C.,  {Johnson} B.~D.,  {Calzetti} D.,  {Dale}
  D.~A.,   {Moustakas} J.,  2011, \mn@doi [\apj] {10.1088/0004-637X/741/2/124},
  \href {http://adsabs.harvard.edu/abs/2011ApJ...741..124H} {741, 124}

\bibitem[\protect\citeauthoryear{{Hayward} \& {Smith}}{{Hayward} \&
  {Smith}}{2015}]{Hayward2015}
{Hayward} C.~C.,  {Smith} D.~J.~B.,  2015, \mn@doi [\mnras]
  {10.1093/mnras/stu2195}, \href
  {http://adsabs.harvard.edu/abs/2015MNRAS.446.1512H} {446, 1512}

\bibitem[\protect\citeauthoryear{{Hayward}, {Kere{\v s}}, {Jonsson},
  {Narayanan}, {Cox}  \& {Hernquist}}{{Hayward} et~al.}{2011}]{Hayward2011}
{Hayward} C.~C.,  {Kere{\v s}} D.,  {Jonsson} P.,  {Narayanan} D.,  {Cox}
  T.~J.,   {Hernquist} L.,  2011, \mn@doi [\apj] {10.1088/0004-637X/743/2/159},
  \href {http://adsabs.harvard.edu/abs/2011ApJ...743..159H} {743, 159}

\bibitem[\protect\citeauthoryear{{Hayward}, {Jonsson}, {Kere{\v s}},
  {Magnelli}, {Hernquist}  \& {Cox}}{{Hayward} et~al.}{2012}]{Hayward2012}
{Hayward} C.~C.,  {Jonsson} P.,  {Kere{\v s}} D.,  {Magnelli} B.,  {Hernquist}
  L.,   {Cox} T.~J.,  2012, \mn@doi [\mnras]
  {10.1111/j.1365-2966.2012.21254.x}, \href
  {http://adsabs.harvard.edu/abs/2012MNRAS.424..951H} {424, 951}

\bibitem[\protect\citeauthoryear{{Hayward}, {Narayanan}, {Kere{\v s}},
  {Jonsson}, {Hopkins}, {Cox}  \& {Hernquist}}{{Hayward}
  et~al.}{2013}]{Hayward2013}
{Hayward} C.~C.,  {Narayanan} D.,  {Kere{\v s}} D.,  {Jonsson} P.,  {Hopkins}
  P.~F.,  {Cox} T.~J.,   {Hernquist} L.,  2013, \mn@doi [\mnras]
  {10.1093/mnras/sts222}, \href
  {http://adsabs.harvard.edu/abs/2013MNRAS.428.2529H} {428, 2529}

\bibitem[\protect\citeauthoryear{{Hayward} et~al.,}{{Hayward}
  et~al.}{2014}]{Hayward2014}
{Hayward} C.~C.,  et~al., 2014, \mn@doi [\mnras] {10.1093/mnras/stu1843}, \href
  {http://adsabs.harvard.edu/abs/2014MNRAS.445.1598H} {445, 1598}

\bibitem[\protect\citeauthoryear{{Hendrix}, {Keppens}  \& {Camps}}{{Hendrix}
  et~al.}{2015}]{Hendrix2015}
{Hendrix} T.,  {Keppens} R.,   {Camps} P.,  2015, \mn@doi [\aap]
  {10.1051/0004-6361/201425498}, \href
  {http://adsabs.harvard.edu/abs/2015A%26A...575A.110H} {575, A110}

\bibitem[\protect\citeauthoryear{{Heyer}, {Carpenter}  \& {Snell}}{{Heyer}
  et~al.}{2001}]{Heyer2001}
{Heyer} M.~H.,  {Carpenter} J.~M.,   {Snell} R.~L.,  2001, \mn@doi [\apj]
  {10.1086/320218}, \href {http://adsabs.harvard.edu/abs/2001ApJ...551..852H}
  {551, 852}

\bibitem[\protect\citeauthoryear{{Hopkins}, {Kere{\v s}}, {O{\~n}orbe},
  {Faucher-Gigu{\`e}re}, {Quataert}, {Murray}  \& {Bullock}}{{Hopkins}
  et~al.}{2014}]{Hopkins2014}
{Hopkins} P.~F.,  {Kere{\v s}} D.,  {O{\~n}orbe} J.,  {Faucher-Gigu{\`e}re}
  C.-A.,  {Quataert} E.,  {Murray} N.,   {Bullock} J.~S.,  2014, \mn@doi
  [\mnras] {10.1093/mnras/stu1738}, \href
  {http://adsabs.harvard.edu/abs/2014MNRAS.445..581H} {445, 581}

\bibitem[\protect\citeauthoryear{{Hughes}, {Cortese}, {Boselli}, {Gavazzi}  \&
  {Davies}}{{Hughes} et~al.}{2013}]{Hughes2013}
{Hughes} T.~M.,  {Cortese} L.,  {Boselli} A.,  {Gavazzi} G.,   {Davies} J.~I.,
  2013, \mn@doi [\aap] {10.1051/0004-6361/201218822}, \href
  {http://adsabs.harvard.edu/abs/2013A%26A...550A.115H} {550, A115}

\bibitem[\protect\citeauthoryear{{James}, {Dunne}, {Eales}  \&
  {Edmunds}}{{James} et~al.}{2002}]{James2002}
{James} A.,  {Dunne} L.,  {Eales} S.,   {Edmunds} M.~G.,  2002, \mn@doi
  [\mnras] {10.1046/j.1365-8711.2002.05660.x}, \href
  {http://adsabs.harvard.edu/abs/2002MNRAS.335..753J} {335, 753}

\bibitem[\protect\citeauthoryear{{Jonsson}, {Groves}  \& {Cox}}{{Jonsson}
  et~al.}{2010}]{Jonsson2010}
{Jonsson} P.,  {Groves} B.~A.,   {Cox} T.~J.,  2010, \mn@doi [\mnras]
  {10.1111/j.1365-2966.2009.16087.x}, \href
  {http://adsabs.harvard.edu/abs/2010MNRAS.403...17J} {403, 17}

\bibitem[\protect\citeauthoryear{{Kennicutt}}{{Kennicutt}}{1998}]{Kennicutt1998}
{Kennicutt} Jr. R.~C.,  1998, \mn@doi [\apj] {10.1086/305588}, \href
  {http://adsabs.harvard.edu/abs/1998ApJ...498..541K} {498, 541}

\bibitem[\protect\citeauthoryear{{Kennicutt} \& {Evans}}{{Kennicutt} \&
  {Evans}}{2012}]{Kennicutt2012}
{Kennicutt} R.~C.,  {Evans} N.~J.,  2012, \mn@doi [\araa]
  {10.1146/annurev-astro-081811-125610}, \href
  {http://adsabs.harvard.edu/abs/2012ARA%26A..50..531K} {50, 531}

\bibitem[\protect\citeauthoryear{Kolmogorov}{Kolmogorov}{1933}]{Kolmogorov1933}
Kolmogorov A.~N.,  1933, Giornale dell'Istituto Italiano degli Attuari, 4, 83

\bibitem[\protect\citeauthoryear{{Laor} \& {Draine}}{{Laor} \&
  {Draine}}{1993}]{Laor1993}
{Laor} A.,  {Draine} B.~T.,  1993, \mn@doi [\apj] {10.1086/172149}, \href
  {http://adsabs.harvard.edu/abs/1993ApJ...402..441L} {402, 441}

\bibitem[\protect\citeauthoryear{{Le Brun}, {McCarthy}, {Schaye}  \&
  {Ponman}}{{Le Brun} et~al.}{2014}]{LeBrun2014}
{Le Brun} A.~M.~C.,  {McCarthy} I.~G.,  {Schaye} J.,   {Ponman} T.~J.,  2014,
  \mn@doi [\mnras] {10.1093/mnras/stu608}, \href
  {http://adsabs.harvard.edu/abs/2014MNRAS.441.1270L} {441, 1270}

\bibitem[\protect\citeauthoryear{{Li} \& {Draine}}{{Li} \&
  {Draine}}{2001}]{Li2001}
{Li} A.,  {Draine} B.~T.,  2001, \mn@doi [\apj] {10.1086/323147}, \href
  {http://adsabs.harvard.edu/abs/2001ApJ...554..778L} {554, 778}

\bibitem[\protect\citeauthoryear{{McAlpine} et~al.,}{{McAlpine}
  et~al.}{2016}]{McAlpine2016}
{McAlpine} S.,  et~al., 2016, \mn@doi [Astronomy and Computing]
  {10.1016/j.ascom.2016.02.004}, \href
  {http://adsabs.harvard.edu/abs/2016A%26C....15...72M} {15, 72}

\bibitem[\protect\citeauthoryear{{McKinnon}, {Torrey}  \&
  {Vogelsberger}}{{McKinnon} et~al.}{2016}]{McKinnon2016}
{McKinnon} R.,  {Torrey} P.,   {Vogelsberger} M.,  2016, \mn@doi [\mnras]
  {10.1093/mnras/stw253}, \href
  {http://adsabs.harvard.edu/abs/2016MNRAS.457.3775M} {457, 3775}

\bibitem[\protect\citeauthoryear{{Morrissey} et~al.,}{{Morrissey}
  et~al.}{2007}]{Morrissey2007}
{Morrissey} P.,  et~al., 2007, \mn@doi [\apjs] {10.1086/520512}, \href
  {http://adsabs.harvard.edu/abs/2007ApJS..173..682M} {173, 682}

\bibitem[\protect\citeauthoryear{{Murphy} et~al.,}{{Murphy}
  et~al.}{2011}]{Murphy2011}
{Murphy} E.~J.,  et~al., 2011, \mn@doi [\apj] {10.1088/0004-637X/737/2/67},
  \href {http://adsabs.harvard.edu/abs/2011ApJ...737...67M} {737, 67}

\bibitem[\protect\citeauthoryear{{Narayanan}, {Hayward}, {Cox}, {Hernquist},
  {Jonsson}, {Younger}  \& {Groves}}{{Narayanan}
  et~al.}{2010a}]{Narayanan2010a}
{Narayanan} D.,  {Hayward} C.~C.,  {Cox} T.~J.,  {Hernquist} L.,  {Jonsson} P.,
   {Younger} J.~D.,   {Groves} B.,  2010a, \mn@doi [\mnras]
  {10.1111/j.1365-2966.2009.15790.x}, \href
  {http://adsabs.harvard.edu/abs/2010MNRAS.401.1613N} {401, 1613}

\bibitem[\protect\citeauthoryear{{Narayanan} et~al.,}{{Narayanan}
  et~al.}{2010b}]{Narayanan2010b}
{Narayanan} D.,  et~al., 2010b, \mn@doi [\mnras]
  {10.1111/j.1365-2966.2010.16997.x}, \href
  {http://adsabs.harvard.edu/abs/2010MNRAS.407.1701N} {407, 1701}

\bibitem[\protect\citeauthoryear{{Nguyen} et~al.,}{{Nguyen}
  et~al.}{2010}]{Nguyen2010}
{Nguyen} H.~T.,  et~al., 2010, \mn@doi [\aap] {10.1051/0004-6361/201014680},
  \href {http://adsabs.harvard.edu/abs/2010A%26A...518L...5N} {518, L5}

\bibitem[\protect\citeauthoryear{{Oppenheimer} et~al.,}{{Oppenheimer}
  et~al.}{2016}]{Oppenheimer2016}
{Oppenheimer} B.~D.,  et~al., 2016, \mn@doi [\mnras] {10.1093/mnras/stw1066},
  \href {http://adsabs.harvard.edu/abs/2016MNRAS.tmp..845O} {}

\bibitem[\protect\citeauthoryear{{Peacock}}{{Peacock}}{1983}]{Peacock1983}
{Peacock} J.~A.,  1983, \mn@doi [\mnras] {10.1093/mnras/202.3.615}, \href
  {http://adsabs.harvard.edu/abs/1983MNRAS.202..615P} {202, 615}

\bibitem[\protect\citeauthoryear{{Poglitsch} et~al.,}{{Poglitsch}
  et~al.}{2010}]{Poglitsch2010}
{Poglitsch} A.,  et~al., 2010, \mn@doi [\aap] {10.1051/0004-6361/201014535},
  \href {http://adsabs.harvard.edu/abs/2010A%26A...518L...2P} {518, L2}

\bibitem[\protect\citeauthoryear{{Press}, {Teukolsky}, {Vetterling}  \&
  {Flannery}}{{Press} et~al.}{2007}]{Press2002}
{Press} W.~H.,  {Teukolsky} S.~A.,  {Vetterling} W.~T.,   {Flannery} B.~P.,
  2007, {Numerical recipes in C++ : the art of scientific computing}, third
  edn.
Cambridge University Press

\bibitem[\protect\citeauthoryear{{Rieke} et~al.,}{{Rieke}
  et~al.}{2004}]{Rieke2004}
{Rieke} G.~H.,  et~al., 2004, \mn@doi [\apjs] {10.1086/422717}, \href
  {http://adsabs.harvard.edu/abs/2004ApJS..154...25R} {154, 25}

\bibitem[\protect\citeauthoryear{{Rieke}, {Alonso-Herrero}, {Weiner},
  {P{\'e}rez-Gonz{\'a}lez}, {Blaylock}, {Donley}  \& {Marcillac}}{{Rieke}
  et~al.}{2009}]{Rieke2009}
{Rieke} G.~H.,  {Alonso-Herrero} A.,  {Weiner} B.~J.,  {P{\'e}rez-Gonz{\'a}lez}
  P.~G.,  {Blaylock} M.,  {Donley} J.~L.,   {Marcillac} D.,  2009, \mn@doi
  [\apj] {10.1088/0004-637X/692/1/556}, \href
  {http://adsabs.harvard.edu/abs/2009ApJ...692..556R} {692, 556}

\bibitem[\protect\citeauthoryear{{Robitaille}, {Churchwell}, {Benjamin},
  {Whitney}, {Wood}, {Babler}  \& {Meade}}{{Robitaille}
  et~al.}{2012}]{Robitaille2012}
{Robitaille} T.~P.,  {Churchwell} E.,  {Benjamin} R.~A.,  {Whitney} B.~A.,
  {Wood} K.,  {Babler} B.~L.,   {Meade} M.~R.,  2012, \mn@doi [\aap]
  {10.1051/0004-6361/201219073}, \href
  {http://adsabs.harvard.edu/abs/2012A%26A...545A..39R} {545, A39}

\bibitem[\protect\citeauthoryear{{Rosas-Guevara} et~al.,}{{Rosas-Guevara}
  et~al.}{2015}]{Rosas-Guevara2015}
{Rosas-Guevara} Y.~M.,  et~al., 2015, \mn@doi [\mnras] {10.1093/mnras/stv2056},
  \href {http://adsabs.harvard.edu/abs/2015MNRAS.454.1038R} {454, 1038}

\bibitem[\protect\citeauthoryear{{Saftly}, {Camps}, {Baes}, {Gordon},
  {Vandewoude}, {Rahimi}  \& {Stalevski}}{{Saftly} et~al.}{2013}]{Saftly2013}
{Saftly} W.,  {Camps} P.,  {Baes} M.,  {Gordon} K.~D.,  {Vandewoude} S.,
  {Rahimi} A.,   {Stalevski} M.,  2013, \mn@doi [\aap]
  {10.1051/0004-6361/201220854}, \href
  {http://adsabs.harvard.edu/abs/2013A%26A...554A..10S} {554, A10}

\bibitem[\protect\citeauthoryear{{Saftly}, {Baes}  \& {Camps}}{{Saftly}
  et~al.}{2014}]{Saftly2014}
{Saftly} W.,  {Baes} M.,   {Camps} P.,  2014, \mn@doi [\aap]
  {10.1051/0004-6361/201322593}, \href
  {http://adsabs.harvard.edu/abs/2014A%26A...561A..77S} {561, A77}

\bibitem[\protect\citeauthoryear{{Saftly}, {Baes}, {De Geyter}, {Camps},
  {Renaud}, {Guedes}  \& {De Looze}}{{Saftly} et~al.}{2015}]{Saftly2015}
{Saftly} W.,  {Baes} M.,  {De Geyter} G.,  {Camps} P.,  {Renaud} F.,  {Guedes}
  J.,   {De Looze} I.,  2015, \mn@doi [\aap] {10.1051/0004-6361/201425445},
  \href {http://adsabs.harvard.edu/abs/2015A%26A...576A..31S} {576, A31}

\bibitem[\protect\citeauthoryear{{Sawala} et~al.,}{{Sawala}
  et~al.}{2016}]{Sawala2016}
{Sawala} T.,  et~al., 2016, \mn@doi [\mnras] {10.1093/mnras/stw145}, \href
  {http://adsabs.harvard.edu/abs/2016MNRAS.457.1931S} {457, 1931}

\bibitem[\protect\citeauthoryear{{Scannapieco}, {Gadotti}, {Jonsson}  \&
  {White}}{{Scannapieco} et~al.}{2010}]{Scannapieco2010}
{Scannapieco} C.,  {Gadotti} D.~A.,  {Jonsson} P.,   {White} S.~D.~M.,  2010,
  \mn@doi [\mnras] {10.1111/j.1745-3933.2010.00900.x}, \href
  {http://adsabs.harvard.edu/abs/2010MNRAS.407L..41S} {407, L41}

\bibitem[\protect\citeauthoryear{{Schaller}, {Dalla Vecchia}, {Schaye},
  {Bower}, {Theuns}, {Crain}, {Furlong}  \& {McCarthy}}{{Schaller}
  et~al.}{2015}]{Schaller2015}
{Schaller} M.,  {Dalla Vecchia} C.,  {Schaye} J.,  {Bower} R.~G.,  {Theuns} T.,
   {Crain} R.~A.,  {Furlong} M.,   {McCarthy} I.~G.,  2015, \mn@doi [\mnras]
  {10.1093/mnras/stv2169}, \href
  {http://adsabs.harvard.edu/abs/2015MNRAS.454.2277S} {454, 2277}

\bibitem[\protect\citeauthoryear{{Schaye}}{{Schaye}}{2004}]{Schaye2004}
{Schaye} J.,  2004, \mn@doi [\apj] {10.1086/421232}, \href
  {http://adsabs.harvard.edu/abs/2004ApJ...609..667S} {609, 667}

\bibitem[\protect\citeauthoryear{{Schaye} \& {Dalla Vecchia}}{{Schaye} \&
  {Dalla Vecchia}}{2008}]{Schaye2008}
{Schaye} J.,  {Dalla Vecchia} C.,  2008, \mn@doi [\mnras]
  {10.1111/j.1365-2966.2007.12639.x}, \href
  {http://adsabs.harvard.edu/abs/2008MNRAS.383.1210S} {383, 1210}

\bibitem[\protect\citeauthoryear{{Schaye} et~al.,}{{Schaye}
  et~al.}{2010}]{Schaye2010}
{Schaye} J.,  et~al., 2010, \mn@doi [\mnras]
  {10.1111/j.1365-2966.2009.16029.x}, \href
  {http://adsabs.harvard.edu/abs/2010MNRAS.402.1536S} {402, 1536}

\bibitem[\protect\citeauthoryear{{Schaye} et~al.,}{{Schaye}
  et~al.}{2015}]{Schaye2015}
{Schaye} J.,  et~al., 2015, \mn@doi [\mnras] {10.1093/mnras/stu2058}, \href
  {http://adsabs.harvard.edu/abs/2015MNRAS.446..521S} {446, 521}

\bibitem[\protect\citeauthoryear{{Schiminovich} et~al.,}{{Schiminovich}
  et~al.}{2007}]{Schiminovich2007}
{Schiminovich} D.,  et~al., 2007, \mn@doi [\apjs] {10.1086/524659}, \href
  {http://adsabs.harvard.edu/abs/2007ApJS..173..315S} {173, 315}

\bibitem[\protect\citeauthoryear{{Skrutskie} et~al.,}{{Skrutskie}
  et~al.}{2006}]{Skrutskie2006}
{Skrutskie} M.~F.,  et~al., 2006, \mn@doi [\aj] {10.1086/498708}, \href
  {http://adsabs.harvard.edu/abs/2006AJ....131.1163S} {131, 1163}

\bibitem[\protect\citeauthoryear{Smirnov}{Smirnov}{1948}]{Smirnov1948}
Smirnov N.,  1948, \mn@doi [Ann. Math. Statist.] {10.1214/aoms/1177730256}, 19,
  279

\bibitem[\protect\citeauthoryear{{Springel}}{{Springel}}{2005}]{Springel2005b}
{Springel} V.,  2005, \mn@doi [\mnras] {10.1111/j.1365-2966.2005.09655.x},
  \href {http://adsabs.harvard.edu/abs/2005MNRAS.364.1105S} {364, 1105}

\bibitem[\protect\citeauthoryear{{Springel}, {White}, {Tormen}  \&
  {Kauffmann}}{{Springel} et~al.}{2001}]{Springel2001}
{Springel} V.,  {White} S.~D.~M.,  {Tormen} G.,   {Kauffmann} G.,  2001,
  \mn@doi [\mnras] {10.1046/j.1365-8711.2001.04912.x}, \href
  {http://adsabs.harvard.edu/abs/2001MNRAS.328..726S} {328, 726}

\bibitem[\protect\citeauthoryear{{Springel}, {Di Matteo}  \&
  {Hernquist}}{{Springel} et~al.}{2005}]{Springel2005a}
{Springel} V.,  {Di Matteo} T.,   {Hernquist} L.,  2005, \mn@doi [\mnras]
  {10.1111/j.1365-2966.2005.09238.x}, \href
  {http://adsabs.harvard.edu/abs/2005MNRAS.361..776S} {361, 776}

\bibitem[\protect\citeauthoryear{{Stalevski}, {Fritz}, {Baes}, {Nakos}  \&
  {Popovi{\'c}}}{{Stalevski} et~al.}{2012}]{Stalevski2012}
{Stalevski} M.,  {Fritz} J.,  {Baes} M.,  {Nakos} T.,   {Popovi{\'c}} L.~{\v
  C}.,  2012, \mn@doi [\mnras] {10.1111/j.1365-2966.2011.19775.x}, \href
  {http://adsabs.harvard.edu/abs/2012MNRAS.420.2756S} {420, 2756}

\bibitem[\protect\citeauthoryear{{Steinacker}, {Baes}  \&
  {Gordon}}{{Steinacker} et~al.}{2013}]{Steinacker2013}
{Steinacker} J.,  {Baes} M.,   {Gordon} K.~D.,  2013, \mn@doi [\araa]
  {10.1146/annurev-astro-082812-141042}, \href
  {http://adsabs.harvard.edu/abs/2013ARA%26A..51...63S} {51, 63}

\bibitem[\protect\citeauthoryear{{Taylor}, {Franx}, {Glazebrook}, {Brinchmann},
  {van der Wel}  \& {van Dokkum}}{{Taylor} et~al.}{2010}]{Taylor2010}
{Taylor} E.~N.,  {Franx} M.,  {Glazebrook} K.,  {Brinchmann} J.,  {van der Wel}
  A.,   {van Dokkum} P.~G.,  2010, \mn@doi [\apj]
  {10.1088/0004-637X/720/1/723}, \href
  {http://adsabs.harvard.edu/abs/2010ApJ...720..723T} {720, 723}

\bibitem[\protect\citeauthoryear{{Taylor} et~al.,}{{Taylor}
  et~al.}{2011}]{Taylor2011}
{Taylor} E.~N.,  et~al., 2011, \mn@doi [\mnras]
  {10.1111/j.1365-2966.2011.19536.x}, \href
  {http://adsabs.harvard.edu/abs/2011MNRAS.418.1587T} {418, 1587}

\bibitem[\protect\citeauthoryear{{Tokunaga} \& {Vacca}}{{Tokunaga} \&
  {Vacca}}{2005}]{Tokunaga2005}
{Tokunaga} A.~T.,  {Vacca} W.~D.,  2005, \mn@doi [\pasp] {10.1086/499029},
  \href {http://adsabs.harvard.edu/abs/2005PASP..117.1459T} {117, 1459}

\bibitem[\protect\citeauthoryear{{Torrey} et~al.,}{{Torrey}
  et~al.}{2015}]{Torrey2015}
{Torrey} P.,  et~al., 2015, \mn@doi [\mnras] {10.1093/mnras/stu2592}, \href
  {http://adsabs.harvard.edu/abs/2015MNRAS.447.2753T} {447, 2753}

\bibitem[\protect\citeauthoryear{{Trayford} et~al.,}{{Trayford}
  et~al.}{2015}]{Trayford2015}
{Trayford} J.~W.,  et~al., 2015, \mn@doi [\mnras] {10.1093/mnras/stv1461},
  \href {http://adsabs.harvard.edu/abs/2015MNRAS.452.2879T} {452, 2879}

\bibitem[\protect\citeauthoryear{{Trayford et al.}}{{Trayford et
  al.}}{prep}]{Trayford2016}
{Trayford et al.} {in prep}, \mnras

\bibitem[\protect\citeauthoryear{{Tremonti} et~al.,}{{Tremonti}
  et~al.}{2004}]{Tremonti2004}
{Tremonti} C.~A.,  et~al., 2004, \mn@doi [\apj] {10.1086/423264}, \href
  {http://adsabs.harvard.edu/abs/2004ApJ...613..898T} {613, 898}

\bibitem[\protect\citeauthoryear{{Viaene} et~al.,}{{Viaene}
  et~al.}{2016}]{Viaene2016a}
{Viaene} S.,  et~al., 2016, \mn@doi [\aap] {10.1051/0004-6361/201527586}, \href
  {http://adsabs.harvard.edu/abs/2016A%26A...586A..13V} {586, A13}

\bibitem[\protect\citeauthoryear{{Vogelsberger} et~al.,}{{Vogelsberger}
  et~al.}{2014}]{Vogelsberger2014}
{Vogelsberger} M.,  et~al., 2014, \mn@doi [\mnras] {10.1093/mnras/stu1536},
  \href {http://adsabs.harvard.edu/abs/2014MNRAS.444.1518V} {444, 1518}

\bibitem[\protect\citeauthoryear{{Wang}, {Dutton}, {Stinson}, {Macci{\`o}},
  {Penzo}, {Kang}, {Keller}  \& {Wadsley}}{{Wang} et~al.}{2015}]{Wang2015}
{Wang} L.,  {Dutton} A.~A.,  {Stinson} G.~S.,  {Macci{\`o}} A.~V.,  {Penzo} C.,
   {Kang} X.,  {Keller} B.~W.,   {Wadsley} J.,  2015, \mn@doi [\mnras]
  {10.1093/mnras/stv1937}, \href
  {http://adsabs.harvard.edu/abs/2015MNRAS.454...83W} {454, 83}

\bibitem[\protect\citeauthoryear{{Whitney}}{{Whitney}}{2011}]{Whitney2011}
{Whitney} B.~A.,  2011, Bulletin of the Astronomical Society of India, \href
  {http://adsabs.harvard.edu/abs/2011BASI...39..101W} {39, 101}

\bibitem[\protect\citeauthoryear{{Wiersma}, {Schaye}  \& {Smith}}{{Wiersma}
  et~al.}{2009a}]{Wiersma2009a}
{Wiersma} R.~P.~C.,  {Schaye} J.,   {Smith} B.~D.,  2009a, \mn@doi [\mnras]
  {10.1111/j.1365-2966.2008.14191.x}, \href
  {http://adsabs.harvard.edu/abs/2009MNRAS.393...99W} {393, 99}

\bibitem[\protect\citeauthoryear{{Wiersma}, {Schaye}, {Theuns}, {Dalla Vecchia}
   \& {Tornatore}}{{Wiersma} et~al.}{2009b}]{Wiersma2009b}
{Wiersma} R.~P.~C.,  {Schaye} J.,  {Theuns} T.,  {Dalla Vecchia} C.,
  {Tornatore} L.,  2009b, \mn@doi [\mnras] {10.1111/j.1365-2966.2009.15331.x},
  \href {http://adsabs.harvard.edu/abs/2009MNRAS.399..574W} {399, 574}

\bibitem[\protect\citeauthoryear{{Wright} et~al.,}{{Wright}
  et~al.}{2010}]{Wright2010}
{Wright} E.~L.,  et~al., 2010, \mn@doi [\aj] {10.1088/0004-6256/140/6/1868},
  \href {http://adsabs.harvard.edu/abs/2010AJ....140.1868W} {140, 1868}

\bibitem[\protect\citeauthoryear{{Zafar} \& {Watson}}{{Zafar} \&
  {Watson}}{2013}]{Zafar2013}
{Zafar} T.,  {Watson} D.,  2013, \mn@doi [\aap] {10.1051/0004-6361/201321413},
  \href {http://adsabs.harvard.edu/abs/2013A%26A...560A..26Z} {560, A26}

\bibitem[\protect\citeauthoryear{{Zahid}, {Dima}, {Kudritzki}, {Kewley},
  {Geller}, {Hwang}, {Silverman}  \& {Kashino}}{{Zahid}
  et~al.}{2014}]{Zahid2014}
{Zahid} H.~J.,  {Dima} G.~I.,  {Kudritzki} R.-P.,  {Kewley} L.~J.,  {Geller}
  M.~J.,  {Hwang} H.~S.,  {Silverman} J.~D.,   {Kashino} D.,  2014, \mn@doi
  [\apj] {10.1088/0004-637X/791/2/130}, \href
  {http://adsabs.harvard.edu/abs/2014ApJ...791..130Z} {791, 130}

\bibitem[\protect\citeauthoryear{{Zibetti}, {Charlot}  \& {Rix}}{{Zibetti}
  et~al.}{2009}]{Zibetti2009}
{Zibetti} S.,  {Charlot} S.,   {Rix} H.-W.,  2009, \mn@doi [\mnras]
  {10.1111/j.1365-2966.2009.15528.x}, \href
  {http://adsabs.harvard.edu/abs/2009MNRAS.400.1181Z} {400, 1181}

\bibitem[\protect\citeauthoryear{{Zubko}, {Dwek}  \& {Arendt}}{{Zubko}
  et~al.}{2004}]{Zubko2004}
{Zubko} V.,  {Dwek} E.,   {Arendt} R.~G.,  2004, \mn@doi [\apjs]
  {10.1086/382351}, \href {http://adsabs.harvard.edu/abs/2004ApJS..152..211Z}
  {152, 211}

\bibitem[\protect\citeauthoryear{{di Serego Alighieri} et~al.,}{{di Serego
  Alighieri} et~al.}{2013}]{diSeregoAlighieri2013}
{di Serego Alighieri} S.,  et~al., 2013, \mn@doi [\aap]
  {10.1051/0004-6361/201220551}, \href
  {http://adsabs.harvard.edu/abs/2013A%26A...552A...8D} {552, A8}

\makeatother
\end{thebibliography}


\appendix

\section{Simulated broad-band photometry}
\label{Photometry.sec}

This appendix summarises the formulae used to convolve a simulated SED with an actual instrument's response curve to obtain a band-integrated flux. We essentially follow the treatments given on Ivan K. Baldry's web page\footnote{\url{http://www.astro.ljmu.ac.uk/~ikb/research/mags-fluxes/}} and in the appendix of \citet{Tokunaga2005}. The procedure depends on whether the instrument counts photons or measures energy.

\subsection{Photon counters}

For a photon counting detector, the number of photons detected per unit time, $t$, and per unit area, $S$, from a source with an intrinsic spectral energy distribution $F_\lambda(\lambda)$, or equivalently $F_\nu(\nu)$, by a photon counter with total system response $R(\lambda)$, or $R(\nu)$, can be written as
\begin{equation}
\frac{\mathrm{d}N_\mathrm{p}}{\mathrm{d}t\,\mathrm{d}S} = \int\frac{\lambda}{hc} F_\lambda(\lambda)R(\lambda) \,\mathrm{d}\lambda
                     = \int\frac{1}{h\nu} F_\nu(\nu)R(\nu) \,\mathrm{d}\nu.
\label{photoncount.eq}
\end{equation}
We define the source's mean intrinsic flux $\left<F_\lambda\right>$ and $\left<F_\nu\right>$ through 
\begin{equation}
\frac{\mathrm{d}N_\mathrm{p}}{\mathrm{d}t\,\mathrm{d}S} = \left<F_\lambda\right> \int\frac{\lambda}{hc} R(\lambda) \,\mathrm{d}\lambda
                     =  \left<F_\nu\right> \int\frac{1}{h\nu} R(\nu) \,\mathrm{d}\nu,
\label{meanfluxdef.eq}
\end{equation}
and we define the pivot wavelength $\lambda_\mathrm{pivot}$ and frequency $\nu_\mathrm{pivot}$ connecting these two flux representations through
\begin{equation}
\left<F_\nu\right> = \left<F_\lambda\right> \frac{\lambda_\mathrm{pivot}^2}{c}
\quad\mathrm{and}\quad
\left<F_\lambda\right> = \left<F_\nu\right> \frac{\nu_\mathrm{pivot}^2}{c}.
\label{pivotdef.eq}
\end{equation}
Combining Eqs.~(\ref{photoncount.eq}) and (\ref{meanfluxdef.eq}) yields 
\begin{equation}
\left<F_\lambda\right> = \frac{ \int\lambda F_\lambda(\lambda)R(\lambda) \,\mathrm{d}\lambda } { \int\lambda R(\lambda) \,\mathrm{d}\lambda } 
\quad\mathrm{and}\quad
\left<F_\nu\right> = \frac{ \int F_\nu(\nu)R(\nu) \,\mathrm{d}\nu/\nu } { \int R(\nu) \,\mathrm{d}\nu/\nu }.
\label{meanflux.eq}
\end{equation}
Substituting $\mathrm{d}\nu/\nu=\mathrm{d}\lambda/\lambda$ in Eq.~(\ref{meanfluxdef.eq}) and combining with Eq.~(\ref{pivotdef.eq}) yields
\begin{equation}
\lambda_\mathrm{pivot} = \sqrt{ \frac{ \int\lambda R(\lambda) \,\mathrm{d}\lambda } {  \int R(\lambda) \,\mathrm{d}\lambda/\lambda } }
\quad\mathrm{and}\quad
\nu_\mathrm{pivot} = \sqrt{ \frac{ \int R(\nu) \,\mathrm{d}\nu/\nu } {  \int R(\nu) \,\mathrm{d}\nu/\nu^3 } }.
\label{pivot.eq}
\end{equation}

\subsection{Bolometers}

For an energy measuring device, the quantities $\lambda R(\lambda)/hc$ and $R(\nu)/h\nu$ in the above analysis must be replaced by the total system transmission $T(\lambda)$ or $T(\nu)$. Eqs.~(\ref{meanflux.eq}) and (\ref{pivot.eq}) then become
\begin{equation}
\left<F_\lambda\right> = \frac{ \int F_\lambda(\lambda)T(\lambda) \,\mathrm{d}\lambda } { \int T(\lambda) \,\mathrm{d}\lambda } 
\quad\mathrm{and}\quad
\left<F_\nu\right> = \frac{ \int F_\nu(\nu)T(\nu) \,\mathrm{d}\nu } { \int T(\nu) \,\mathrm{d}\nu }
\label{meanfluxbol.eq}
\end{equation}
and
\begin{equation}
\lambda_\mathrm{pivot} = \sqrt{ \frac{ \int T(\lambda) \,\mathrm{d}\lambda } {  \int T(\lambda) \,\mathrm{d}\lambda/\lambda^2 } }
\quad\mathrm{and}\quad
\nu_\mathrm{pivot} = \sqrt{ \frac{ \int T(\nu) \,\mathrm{d}\nu } {  \int T(\nu) \,\mathrm{d}\nu/\nu^2 } }.
\label{pivotbol.eq}
\end{equation}


\bsp	
\label{lastpage}
\end{document}